\documentclass[preprint2]{proto}
\usepackage{times}
\newcommand{\refs}{\par\noindent\hangindent=1pc\hangafter=1}
\newcommand \lsol{L$_{\odot}$}
\newcommand \msol{M$_{\odot}$}
\newcommand \mearth{M$_{\oplus}$}

\begin{document}

\title{\textbf{\LARGE Extra-Solar Kuiper Belt Dust Disks}}

\author {\textbf{\large Amaya Moro-Mart\'{\i}n}}
\affil{\small\em Princeton University}

\author {\textbf{\large Mark C. Wyatt}}
\affil{\small\em University of Cambridge}

\author {\textbf{\large Renu Malhotra and David E. Trilling}}
\affil{\small\em University of Arizona}

\begin{abstract}
\baselineskip = 11pt
\leftskip = 0.65in 
\rightskip = 0.65in
\parindent=1pc
{\small 
The dust disks observed around mature stars are evidence that plantesimals are present in 
these systems on spatial scales that are similar to that of the asteroids and the KBOs in the Solar System. 
These dust disks (a.k.a. ``debris disks'') present a wide range of sizes, morphologies and properties. It is 
inferred that their dust mass declines with time as the dust-producing 
planetesimals get depleted, and that this decline can be punctuated by large spikes that are
produced as a result of individual collisional events. The lack of solid state features indicate that, generally, 
the dust in these disks have sizes $\gtrsim$10~$\mu$m, but exceptionally, strong silicate 
features in some disks suggest the presence of large quantities of small grains, thought to be the result of recent collisions. 
Spatially resolved observations of debris disks show a diversity of structural features, such as inner cavities, 
warps, offsets, brightness asymmetries, spirals, rings and clumps. There is growing evidence that, in some cases, 
these structures are the result of the dynamical perturbations of a massive planet. 
Our Solar System also harbors a debris disk and some of its properties resemble those of extra-solar 
debris disks. From the cratering record, we can infer that its 
dust mass has decayed with time, and that there was at least one major ``spike'' in the 
past during the Late Heavy Bombardment. This offers a unique opportunity to use 
extra-solar debris disks to shed some light in how the Solar System might have 
looked in the past. Similarly, our knowledge of the Solar System is influencing our
understanding of the types of processes which might be at play in the extra-solar 
debris disks. 
\\~\\~\\~}

\end{abstract}  

\section{\textbf{INTRODUCTION}}
During the last two decades, space-based infrared observations, first with $\it{IRAS}$ 
and then with $\it{ISO}$ and $\it{Spitzer}$, have shown that main sequence stars are 
commonly surrounded by dust disks (a.k.a. debris disks), some of which extend to 100s 
of AU from the central star. With the recent $\it{Spitzer}$ observations, the number of debris 
disks known to date is approaching 100, of which 11 are spatially resolved. 

Dust particles are affected by radiation pressure, 
Poynting-Robertson and stellar wind drag, mutual collisions and collisions with 
interstellar grains. All these processes contribute to make the lifetime of the dust 
particles significantly shorter than the age of the star. Therefore, it was 
realized early on that this dust could not be primordial, i.e. part of the original molecular cloud 
where the star once formed, but it had to be a second generation of dust, likely replenished 
by a reservoir of (undetected) dust-producing planetesimals 
like the asteroids, comets and Kuiper Belt Objects (KBOs) in our solar system ({\em Backman and Paresce}, 1993). 
This represented a major leap in the 
search for other planetary systems: by 1983, a decade before extra-solar planets were 
discovered, $\it{IRAS}$ observations proved that there is planetary material surrounding 
nearby stars ({\em Aumann et al.}, 1984). 

How do the extra-solar debris disks compare with our own Solar System? The existence
of an inner planetary dust complex has long been known from observations of 
zodiacal light ({\em Cassini}, 1683). 
In the inner Solar System, dust is produced by debris from Jupiter family short 
period comets and asteroids ({\em Liou, Dermott and Xu}, 1995; {\em Dermott et al.}, 1994). The
scattering of sunlight by these grains gives rise to the zodiacal 
light and its thermal emission dominates the night sky 
between 5 $\mu$m and 500 $\mu$m. This thermal emission dust was observed 
by the $\it{IRAS}$ and $\it{COBE}$ space telescopes, 
and the interplanetary dust particles (IDPs) were detected {\em in situ} by dust 
detectors on Pioneer 10 and 11, Voyager, Galileo and Ulyses spacecrafts. Its 
fractional luminosity is estimated to be L$_{dust}$/L$_*$$\sim$10$^{-8}$--10$^{-7}$
({\em Dermott et al.}, 2002). 
In the outer Solar System, significant dust production is expected from the 
mutual collisions of KBOs and collisions with interstellar grains ({\em Backman and Paresce}, 1993; 
{\em Stern}, 1996; {\em Yamamoto and Mukai}, 1998). 
The thermal emission of the outer Solar System dust is overwhelmed by the much stronger signal from the inner zodiacal 
cloud (so KB dust is not seen in the $\it{IRAS}$ and $\it{COBE}$ infrared maps). However,  
evidence of its existence comes from the Pioneer 10 and 11 dust collision events 
measured beyond the orbit of Saturn ({\em Landgraf et al.}, 2002). Extrapolating from the 
size distribution of KBOs, its fractional luminosity is estimated to be 
L$_{dust}$/L$_*$$\sim$10$^{-7}$--10$^{-6}$ ({\em Stern}, 1996). 

In this chapter we describe the debris disk phenomenon: how debris 
disks originate ($\S$2); how they evolve in time ($\S$3);
what are they made of ($\S$4); whether or not they are related to the presence
of close-in planets ($\S$5); and how planets can affect their structure ($\S$6). 
We then discuss how debris disks compare to the Solar System's dust disk in the 
present and in the past ($\S$7), and finish with a discussion of the prospects for 
the future of debris disk studies ($\S$8). In summary, the goal of the chapter is to 
review how debris disks can help us place our Solar System into context within extra-solar planetary systems. 

\bigskip

\centerline{\textbf{ 2. FROM PRIMORDIAL TO DEBRIS DISKS}}
\bigskip

Stars form from the collapse of dense regions of molecular clouds, and a 
natural by-product of this process is the formation of a circumstellar disk
({\em Shu, Adams and Lizano}, 1987; {\em Hartmann}, 2000).
Observations show that young stars with masses below $\sim$4~\msol~ down to 
brown dwarfs and planetary-mass objects have disks, 
while disks around more massive stars are more elusive, due to fast disk 
dissipation and observational difficulties as they tend to be highly embedded and typically 
very distant objects. 
Disk masses are estimated to be in the range 0.003 \msol--0.3 \msol, showing a large spread even for
stars with similar properties ({\em Natta}, 2004 and references therein).
For one solar mass stars, disk masses are 0.01 \msol --0.10 \msol~
({\em Hartmann}, 2000 and references therein).
With regard to the disks sizes, there is evidence for gas
on scales from 10 AU to 800 AU ({\em Simon, Dutrey and Gilloteau}, 2000). Both the disk masses and scales are 
comparable to the minimum mass solar nebula, $\sim$0.015 \msol. This is 
the total mass of solar composition material needed to produced the observed condensed 
material in the Solar System planets ($\sim$50 \mearth; {\em Hayashi}, 1981; {\em Weidenschilling}, 1977).

Eventually, infall to the disk stops and the disk becomes depleted in mass:
most of the disk mass is accreted onto the central star; some material may be 
blown away by stellar wind ablation or by photo-evaporation by high-energy 
stellar photons, or stripped away by interactions with passing stars; 
the material that is left behind might coagulate or 
accrete to form planets (only $\sim$10\% of the solar nebula gas is accreted
into the giant planet's atmospheres). 
After $\sim$10$^{7}$ years, most of the primordial gas and dust have disappeared 
(see e.g. {\em Hollenbach et al.}, 2005; {\em Pascucci et al.}, 2006), setting an important time 
constraint for giant planet formation models. 

However, many main sequence stars older than $\sim$10$^{7}$ years still 
show evidence of dust. The timescale of dust grain removal due to 
radiation pressure is of the order of an orbital period, while the
Poynting-Robertson (P-R) drag lifetime of a dust grain located at a distance R
is given by 
$$ t_{PR} = 710 ({b \over \mu m}) ({\rho \over g/cm^3}) ({R \over AU})^2 
({L_\odot \over L_*}) {1 \over 1+albedo} ~yr, $$ 
where $\it{b}$ and $\rho$ are the grain radius and density, respectively
({\em Burns, Lamy and Soter}, 1979 and {\em Backman and Paresce}, 1993). Grains can also be destroyed 
by mutual grain collisions, with a collisional lifetime of
$$  t_{col} = 
1.26 \times 10^4 ({R \over AU})^{3/2} ({M_\odot \over M_*})^{1/2} 
({10^{-5} \over L_{dust}/L_*}) yr $$
({\em Backman and Paresce}, 1993). Because all the above timescales are generally much shorter than 
the age of the disk, it is inferred that the observed dust is not primordial 
but is likely produced by a reservoir of undetected kilometer-sized 
planetesimals producing dust by mutual collisions or by evaporation of comets 
scattered close to the star ({\em Backman and Paresce}, 1993).

\bigskip
\noindent
\textbf{2.1 Debris Disk Diversity as a Result of the Starting}

\textbf{~~Conditions}
\bigskip

At any particular age, observations show a great diversity of debris disks 
surrounding similar type stars (see $\S$3.1 and {\em Andrews and Williams}, 2005). 
This may be due to the following 
factors that can influence the disks at different stages during their evolution: 
a) different initial masses and sizes, 
caused by variations in the angular momentum of the collapsing protostellar cloud; 
b) different external environments, causing variations in the dispersal time scales 
of the outer primordial disks, and therefore strongly affecting the formation of planets
and planetesimals in the outer regions; and
c) different planetary configurations, affecting the populations and velocity dispersions 
of the dust-producing planetesimals. 

For example, the formation environment can have an important effect on the disk size and 
its survival. If the star is born in a sparsely populated Taurus-like association, the
possibility of having a close encounter with another star that could truncate the outer
protoplanetary disk is very small. In this environment, the probability of having a nearby massive star 
is also small, so photoevaporation does not play an important role in 
shaping the disk, and neither does the effect of explosions of nearby supernovae
({\em Hollenbach and Adams}, 2004). However, if the star is born in a densely populated OB association, 
the high density of stars results in a high probability of close encounters that could
truncate the outer protoplanetary disk. In addition, nearby massive stars and supernovae 
explosions are likely to be present, affecting the size of the disk by 
photoevaporation, a process in which the heated gas from the outer disk evaporates into 
interstellar space, dragging along dust particles smaller than a critical size of
0.1 cm --1 cm before they have time to coagulate into larger bodies. 
Dust coagulation to this critical size takes $\sim$ 10$^{5}$ yr--10$^{6}$ yr 
at 30 AU--100 AU ({\em Hollenbach and Adams}, 2004) , 
and therefore occurs rapidly enough for KB formation to take place inside 100 AU, 
even around low mass stars in OB associations like the Trapezium in Orion. 
However, in Trapezium-like conditions ({\em Hillenbrand and Hartman}, 1998), 
where stars form within groups/clusters containing $>$100 
members , at larger distances from the star photoevaporation takes place on a faster time scale 
than coagulation, and the dust is carried away by the evaporating gas causing a sharp cutoff in 
the formation of planetesimals beyond $\sim$ 100(M$_{star}$/1\msol) AU, and therefore 
suppressing the production of debris dust ({\em Hollenbach and Adams}, 2004 and references therein). 
Debris disks can present a wide range of sizes because the distance at which photoevaporation 
takes place on a faster time scale than coagulation depends not only on the 
mass of the central star, but also on the initial disk mass and the mass and proximity of the most 
massive star in the group/cluster. 

It is thought that the sun formed in an OB association: 
meteorites show clear evidence that isotopes with short lifetimes ($<$10$^{5}$ yr) 
were present in the solar nebula, which indicates that a nearby supernova introduced them immediately 
before the dust coagulated into larger solids ({\em Cameron and Truran}, 1977; 
{\em Tachibana et al.}, 2006); and in addition, it has been suggested that the edge of 
the KB may be due to the dynamical interaction with a passing star ({\em Kobayashi, Ida and Tanaka}, 2005), 
indicating that the sun may have been born in a high density stellar environment.
In contrast, kinematic studies show that the majority of the nearby spatially resolved debris disks 
formed in loosely populated Taurus-like associations (see e.g. {\em Song, Zuckerman and Bessell}, 2003). 

Debris disks found around field stars may be 
intrinsically different than those found around stars that once belonged to densely 
populated clusters, and one needs be cautious of the conclusions drawn from comparing
these systems directly, as well as the conclusions drawn from stellar samples that 
include indiscriminately debris disks forming in these two very different environments.

\bigskip

\centerline{\textbf{ 3. DEBRIS DISKS EVOLUTION AND FREQUENCY}}
\bigskip

The study of debris disk evolution, i.e. the dependency on stellar age of the amount of dust
around a main sequence star, is of critical importance in the understanding of the timescales 
for the formation and evolution of planetary systems, as the dust production rate is thought 
to be higher during the late stages of planet formation, when planetesimals are colliding 
frequently, than later on, when mature planetary systems are in place, planet formation is 
complete and the planets are not undergoing migration. 
Because it is obviously not possible to observe in real time the evolution of a particular system
during Myr--Gyr, the study of debris disk evolution is based on the observations of a large number
of stars with different ages, with the goal of determining how the amount of excess emission (related to the dust 
mass) and the probability of finding an excess depend on stellar age. 
The assumption is that all the disks will evolve in a similar way (but see caveats in $\S$2.1).

The age-dependency of the dust emission (a.k.a ``excess'' 
with respect to the photospheric values) has been elusive until recently. 
The limited sensitivity of $\it{IRAS}$ allowed only the detection
of the brightest and nearest disks, mostly around A stars. In addition, with its 
limited spatial resolution it was not possible to determine whether the infrared excess 
emission was coming from the star (i.e., from a debris disks) or from extended galactic 
cirrus or background galaxies. $\it{ISO}$, with its improvement of a factor
of two in spatial resolution and a factor of 10 in sensitivity over $\it{IRAS}$, 
made a big step forward in the study of debris disks evolution.
However, the $\it{ISO}$ samples were too small to establish any age-dependency
on a sound statistical basis. More recently, the $\it{Spitzer}$/MIPS instrument, with its 
unprecedented sensitivity at far-IR wavelengths (a factor of $\sim$100-1000 better than $\it{IRAS}$,
and at least a factor of 10 in spatial resolution), 
has extended the search of disks around main sequence stars to more tenuous disks and to 
greater distances, providing more homogeneous samples. This is still on-going research but is 
leading to new perspective on debris disk evolution. The following subsections summarize the 
main results so far. 

\bigskip
\noindent
\textbf{3.1 Observations}

\bigskip
\noindent
{\em 3.1.1 A stars}\\
\\
Using $\it{Spitzer}$/MIPS at 24~$\mu$m, {\em Rieke et al.} (2005) carried out a survey 
of 76 A-stars (2.5 \msol) of ages 5 Myr--580 Myr, 
with all the stars detected to 7--$\sigma$ 
relative to their photospheric emission. These observations were 
complemented with archival data from $\it{ISO}$ and $\it{IRAS}$, resulting in a total of 
266 A-stars in the final sample studied. The results show an overall decline in the average 
amount of 24~$\mu$m excess emission. Large excesses (more than a factor
of 2 relative to the photosphere) decline from $\sim$25\% in the youngest age bins to 
only one star ($\sim$1\%) for ages $>$190 Myr; a functional fit to this data suggest 
a $\it{t}$$_{0}$/$\it{t}$ decline, with $\it{t}$$_{0}$=100 Myr--200 Myr. 
Intermediate excesses (factors of 1.25--2) decrease much more slowly and are present
in $\sim$7\% of stars older than several hundred Myr. The persistence of excesses 
beyond 200 Myr rules out a fast 1/$\it{t}$$^{2}$ decay. 
Using a sub-sample of 160 A-stars (including the ones in {\em Rieke et al.}, 2005),  
{\em Su et al.} (2006) confirmed that the 24~$\mu$m excess emission is consistent with a 
$\it{t}$$_{0}$/$\it{t}$ decay, where $\it{t}$$_{0}$$\sim$150 Myr, while the 70~$\mu$m excess (tracing dust 
in the KB region) is consistent with $\it{t}$$_{0}$/$\it{t}$, where $\it{t}$$_{0}$$\gtrsim$400 Myr. 

Even though there is a clear decay of the excess emission with time, {\em Rieke et al.} (2005) 
and {\em Su et al.} (2006) showed that at a given stellar age there are 
at least two orders of magnitude variations in the amount of dust: 
as many as 50\%--60\% of the younger stars ($<$30 Myr) do not show dust 
emission at 24~$\mu$m, while $\sim$ 25\% of disks are still detected at 150 Myr.

\bigskip
\noindent
{\em 3.1.2. FGK stars}\\
\\
For FGK stars, the excess rates at 24 $\mu$m decrease from $\sim$30\%--40\% for ages $<$50 Myr,  
to $\sim$9\% for 100 Myr--200 Myr, and $\sim$1.2\% for ages $>$1 Gyr 
(see Figure 1; {\em Siegler et al.}, 2006; {\em Gorlova et al.}, 2006; {\em Stauffer et al.}, 2005; {\em Beichman et al.}, 
2005a; {\em Kim et al.}, 2005; {\em Bryden et al.}, 2006). At 70 $\mu$m, the excess rate is
10\%--20\% and is fairly constant for a wide range of ages ({\em Bryden et al.}, 2006; {\em Hillenbrand et al.}, in preparation). 
At first sight, it appears that for the older stars warm asteroid belt-like disks are rare (few percent), while 
cold KB-like disks are common (10\%--20\%). However, one needs to keep in mind that 
the sensitivity thresholds at 24 $\mu$m~and 70 $\mu$m~
are different: $\it{Spitzer}$/MIPS is currently able to constrain dust masses at KB-like distances (10 AU--100 AU) 
that are 5--100 times the level of dust in our Solar System, and at AB-like distances (1--10 AU) that are 
1000 times our zodiacal emission ({\em Bryden et al.}, 2006). 
However, spectroscopy observations with $\it{Spitzer}$/IRS are better suited to search for hot dust. 
Preliminary results by {\em Beichman et al.} (2006a) indicated that indeed warm excesses ($<$25~$\mu$m) with luminosities
50--1000 times the zodiacal emission are rare for stars $>$1 Gyr, found around only 1 out of 40 stars and
in agreement with theoretical calculations of disk dispersal by {\em Dominik and Decin} (2003) 
that indicate that the fractional luminosity of the warm dust will generally drop 
below the IRS detectability level after 1 Gyr of evolution. In contrast, 
colder disks with excesses at 30--34$\mu$m, are found around 5 out of 41 stars, 12$\pm$5\%, 
in agreement with {\em Bryden et al.}, (2006). 

Even though the $\it{Spitzer}$/MIPS detection rate of excess emission for FGK stars is lower than 
for A-stars (see Figure 1), this is also a result of a sensitivity threshold: similar
levels of excess emission are more easily detected around 
hotter stars than around colder stars. Accounting for this, the actual frequency of debris disks does 
not seem to be a strong function of stellar type ({\em Siegler et al.}, 2006), 
but it drops to zero for stars later than K1 ({\em Beichman et al.}, 2006b).

As for A-stars, FGK stars also show large variations in the amount of excess emission at a given stellar age at
24 $\mu$m (see Figure 2) and 70 $\mu$m. In addition, {\em Siegler et al.} (2006) found that the upper 
envelope of the ratio of the excess emission over the stellar photosphere at 24 $\mu$m~also decays as 
$\it{t}$$_{0}$/$\it{t}$, with $\it{t}$$_{0}$=100 Myr and ages $>$20 Myr. At younger ages, $<$25 Myr, the decay 
is significantly faster and could trace the fast transition of the disk between primordial and debris stages 
({\em Siegler et al.}, 2006). For colder dust (at 70 $\mu$m), even though there is a general trend to find less 
dust at older ages, the decay time is longer than for warmer dust (at 24 $\mu$m). 

\begin{figure*}
 \epsscale{1.0}
\begin{center}
 \includegraphics[angle=90,scale=0.5]{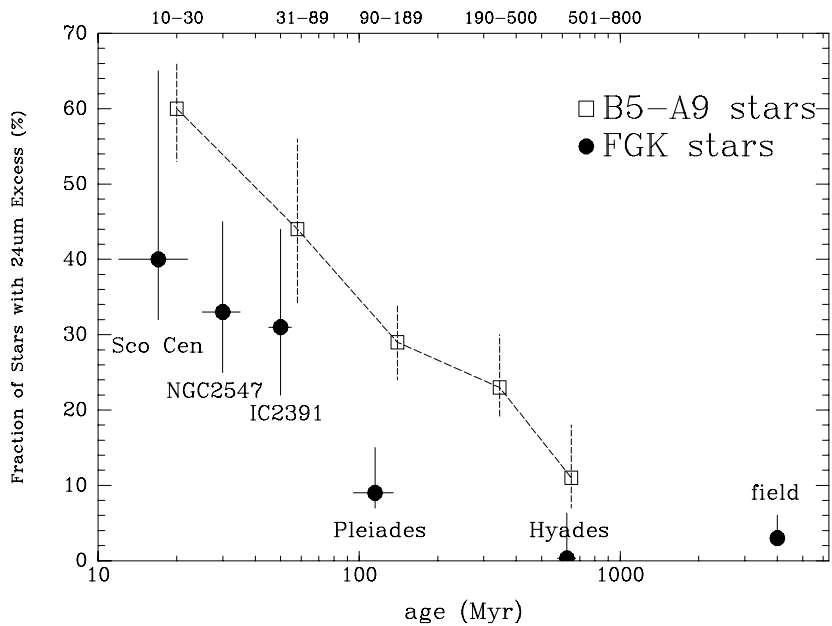}
\end{center}
 \caption{\small 
Fraction of early-type stars (open squares) and FGK stars (circles) with excess emission at 24 $\mu$m
as a function of stellar age. Figure from {\em Siegler et al.} (2006) using data 
from {\em Chen et al.} (2005), {\em Gorlova et al.} (in prep), {\em Stauffer et al.} (2005), {\em Gorlova et al.} (2006),
{\em Cieza, Cochran and Paulson} (2005), {\em Bryden et al.} (2006), {\em Rieke et al.} (2005) and {\em Su et al.} (2006). 
The age bins used in the early-type star survey are shown across the the top horizontal axis. 
Vertical error bars are 1-sigma binomial distribution uncertainties. 
}
\end{figure*}

\begin{figure*}
 \epsscale{1.0}
\begin{center}
 \includegraphics[angle=90,scale=0.5]{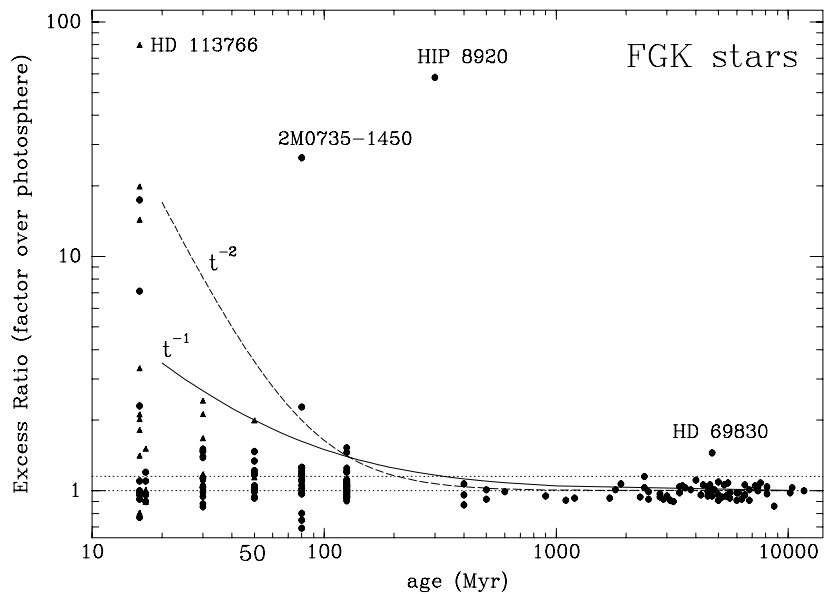}
\end{center}
 \caption{\small Ratio of the 24 $\mu$m~excess emission to the predicted photospheric value for FGK
stars as a function of stellar age. Triangles represent F0--F4 stars and circles represent F5--K7 stars (similar 
to the Sun). Stars aligned vertically belong to clusters or associations. Figure from {\em Siegler et al.} (2006)
using the same data as in Figure 1 and from {\em Gorlova et al.} (2004), {\em Hines et al.} (2006)
and {\em Song et al.} (2005).
}
\end{figure*}

\bigskip
\noindent
\textbf{3.2 Theoretical Predictions}

\bigskip
\noindent
{\em 3.2.1 Inverse-Time Decay}\\
\\
If all the dust is derived from the grinding down of
planetesimals, and assuming the planetesimals are destroyed after one collision, 
and that the number of collisions is proportional to the square of the number of 
planetesimals ($\it{N}$), then d$\it{N}$/d$\it{t}$ $\propto$ -$\it{N}$$^{-2}$ and $\it{N}$ $\propto$ 1/$\it{t}$. 
Therefore, the dust production rate, $\it{R}$$_{prod}$ $\propto$ 
d$\it{N}$/d$\it{t}$ $\propto$ $\it{N}$$^{2}$ $\propto$ 1/$\it{t}$$^{2}$. 
To solve for the amount of dust in the disk in steady state, one needs equate the dust 
production rate to the dust loss rate, $\it{R}$$_{loss}$, and this gives two different 
solutions depending on the number density of the dust in the disk ({\em Dominik and Decin}, 2003):
(1) In the collisionally-dominated disks ($\it{M}$$_{dust}$$\gtrsim$10$^{-3}$~\mearth), 
the dust number density is high and the main dust removal process are grain-grain 
collisions, so that $\it{R}$$_{loss}$ $\propto$ $\it{n}$$^{2}$, where $\it{n}$ is the number of dust grains.  
From $\it{R}$$_{prod}$ = $\it{R}$$_{loss}$, we get that $\it{n}$ $\propto$ 1/$\it{t}$. 
(2) In the radiatively-dominated disks ($\it{M}$$_{dust}$$\lesssim$10$^{-3}$~\mearth), the dust loss rate is 
dominated by Poynting-Robertson drag, and therefore is proportional to the number of particles, 
$\it{R}$$_{loss}$ $\propto$ $\it{n}$, and from $\it{R}$$_{prod}$ = $\it{R}$$_{loss}$, we get that $\it{n}$ 
$\propto$ 1/$\it{t}$$^{2}$. 

The Kuiper Belt disk has little mass and is radiatively dominated. However,
all the debris disks observed so far are significantly more massive than the KB because
the surveys are sensitivity limited. {\em Wyatt} (2005a) estimates that the observed disks 
are generally collisionally dominated, 
so one would expect that the dust emission will evolve as 1/$\it{t}$, in agreement with the 
Spitzer/MIPS observations of debris disks around A and FGK stars.

\bigskip
\noindent
{\em 3.2.2. Episodic Stochastic Collisions}\\
\\
Numerical simulations of the evolution of dust generated from the collision of planetesimals around solar-type
stars by {\em Kenyon and Bromley} (2005) predict that after 1 Myr there is a steady decline of the 24 $\mu$m~excess 
emission, as the dust-producing planetesimals get depleted, a decay that is punctuated by large spikes 
produced by individual collisional events (see Figure 3). 
Therefore, the high degree of debris disk variability observed by Spitzer/MIPS
- seen as spikes in Figure 2 - may be the result of recent collisional events. 
It is thought that these events initiate a collisional cascade 
leading to short-term increases in the density of small grains, which increases the brightness density of the 
disk by an order of magnitude. Because the clearing time of dust in the 24~$\mu$m-emitting zone (10 AU--60 AU)
is $\sim$1 Myr--10 Myr ({\em Dominik and Decin}, 2003; {\em Kenyon and Bromley}, 2004), these individual events could 
dominate the properties of debris disks over Myr timescales ({\em Rieke et al.}, 2005). 
However, there is a discrepancy between these numerical simulations and the observations because
the models do not predict excess ratios larger than two for stars older than 
50 Myr, in disagreement with the existence of two of the outliers in Figure 2 (HIP 8920 and 2M0735-1450). 

\begin{figure*}
 \epsscale{1.0}
 \plotone{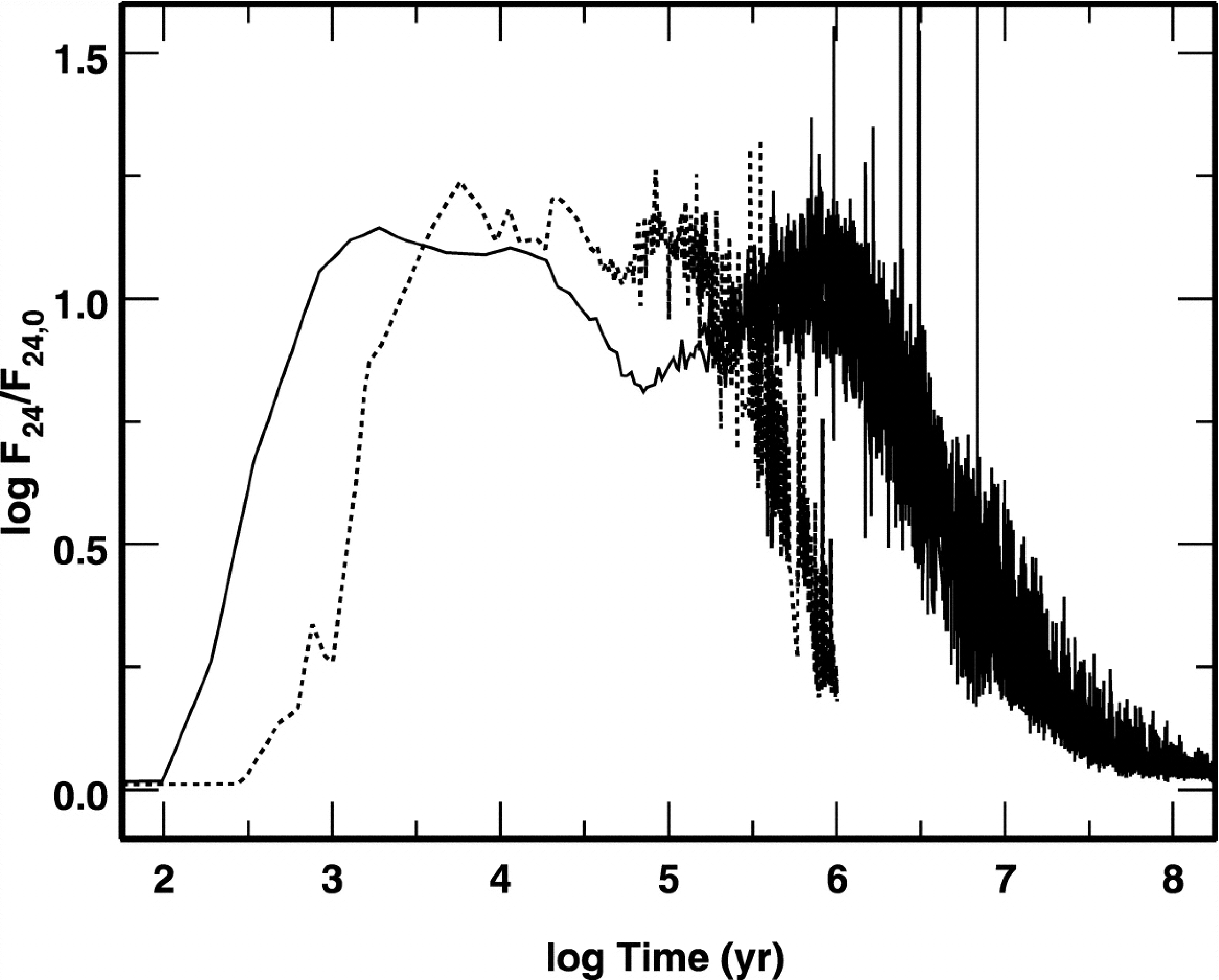}
 \caption{\small Evolution of the 24 $\mu$m excess as a function of time for two planetesimal 
disks extending from 0.68--1.32 AU (dashed line) and 0.4--2 AU (solid line). The central star
is solar type. Excess emission decreases as planetesimals grow into Mars-sized or larger objects 
and collisions become increasingly rare. Figure from {\em Kenyon and Bromley} (2005).
}
\end{figure*}

In addition to the large differences in excess emission found among stars within the same age range
(for both A stars and FGK stars), the presence of large amounts of small grains in systems like HIP 8920 and 
HD 69830 (two of the outliers in Figure 2), Vega, and in a clump in $\beta$-Pic ({\em Telesco et al.}, 2005), 
indicate that recent collisional events have taken place in these systems (see discussion in $\S$4). 
The argument goes as follows: because small grains are removed quickly by radiation pressure, 
the dust production rate needed to account for the observations is very high, implying
a mass loss that could not be sustained during the full age of the system. 
For example, in the Spitzer/MIPS observations of Vega (350 Myr old A star) show that the disk at
24~$\mu$m and 70~$\mu$m extends to distances of 
330 AU and 540 AU from the star, respectively ({\em Su et al.}, 2006), 
far outside the $\sim$80 AU ring of dust seen in the submillimeter ({\em Wilner et al.}, 2002) 
that probably traces the location of the dust-producing planetesimals. 
{\em Su et al.} (2006) suggested that the dust observed in the mid-IR comes from small 
grains that were generated in a recent collisional event that took place in the planetesimals 
belt, and that are being expelled from the system under radiation pressure. This scenario would explain 
the large extent of the disk and the unusually high dust production rate (10$^{15}$ g/s), 
unsustainable for the entire lifetime of Vega. 

\bigskip

\centerline{\textbf{ 4. DEBRIS DISK GRAIN SIZE AND COMPOSITION}}
\bigskip

Most debris disk spectroscopy observations show little or no solid state features, 
indicating that at those stages the dust grains have sizes $\gtrsim$10~$\mu$m 
({\em Jura et al.}, 2004; {\em Spapelfeldt et al.}, 2004), 
much larger than the sub-micron amorphous silicate grains that dominate the dust emission 
in young protoplanetary disks. While for A-stars, the lack of features is explained by
the ejection of dust grains $<$10~$\mu$m by radiation pressure, 
the reason why this is also the case in debris disks around solar-type stars is still 
under debate. However, there are a few debris disks where spectral features have been observed, allowing 
us to set constrains on the particle size and composition. We briefly 
describe three of these systems: $\beta$-Pictoris, for which small quantities 
of silicates have been observed, and HIP 8920 and HD 69830, showing 
very strong silicate features. 

$\beta$-Pictoris is one of the youngest and closest (19 pc) 
stars to Earth harboring a disk.
It is an A5V star (2 \msol) with an estimated age of 12 Myr probably in the 
process of clearing out its protoplanetary disk, as the Solar System did 4 billion 
years ago. 
The disk is likely in the transition between the primordial and debris stages. 
Its dust disk, seen edge-on, extends to 1000 AU (i.e. $\sim$10 times that 
of the solar system) and contains a few lunar masses in grains that are relatively 
large ($>$1$\mu$m), with a large fractional luminosity, L$_{dust}$/L$_{star}$ 
$\sim$ 3$\times$10$^{-3}$. The break in the surface brightness profile of the disk 
indicates that the outer edge of the dust-producing planetesimal belt is at 
$\sim$ 120 AU ({\em Heap et al.}, 2000). Small particles produced by collisions in the 
belt are diffused out by radiation pressure, explaining the power law index of
the brightness profile. On a smaller scale, spatially resolved spectroscopy 
observations indicate that the disk emission is dominated by grains emitting in the 
continuum, with moderate silicate emission features (amorphous and crystalline) seen only 
within 25 AU of the star. This indicates that the ratio of small to large silicate grains 
decreases with distance ({\em Weinberger, Becklin and Zuckerman}, 2003). 
Additional spatially resolved spectroscopy observations by {\em Okamoto et al.} (2004) 
showed that the sub-micron amorphous silicate grains have three peaks in their distribution 
around 6 AU, 16 AU and 30 AU, and their locations possibly trace three belts of dust-producing 
planetesimals. Finally, 
in the innermost system, the gas absorption lines detected toward the star indicate that there is a stable 
gas component that is located at about 1 AU and can be explained by the 
replenishment of gas by evaporating comets near the star, which would also give rise to the 
transient redshifted absorption events observed in the spectra. The frequency of star-grazing
comets needed to explain the observations is several orders of magnitude higher than 
that found in the Solar System (see review in {\em Lagrange, Backman and Artymowicz}, 2000). 

HIP 8920 (one of the outliers in Figure 2) is a 300 Myr old star with a disk that has a high 
surface density of small ($\lesssim$2.5~$\mu$m) dust grains at 1 AU from the star. 
Mid-infrared spectroscopy observations of the dust emission at 8~$\mu$m--13~$\mu$m show a 
very strong silicate feature with broad peaks at 10 and 11~$\mu$m that can be modeled with 
a mixture of amorphous and crystalline silicate grains (pyroxenes and olivines), with sizes of 
0.1~$\mu$m--2.5~$\mu$m. Because HIP 8920 is too old for the dust to be primordial, 
it has been suggested that the anomalous large quantities of small
grains could be the result of a recent collision ({\em Weinberger et al.}, 
priv. comm.).

HD~69830 is a 2 Gyr old K0V star (0.8~\msol, 0.45~\lsol) 
with an excess emission at 8 $\mu$m--35 $\mu$m (60\% over the photosphere at 35~$\mu$m, and 
with fractional luminosity L$_{dust}$/L$_{star}$ $\sim$ 2$\times$10$^{-4}$) 
that shows strong silicate features remarkably similar to the ones in the comet  C/1995 O1 (a.k.a Hale-Bopp -- see Figure 4 from {\em Beichman et al.}, 2005b). 
The spectral features are identified as arising from mostly crystalline olivine 
(including fosterite), and 
a small component of crystalline pyroxene (including enstatite), both of which are also found in interplanetary dust 
particles and meteorite inclusions ({\em Yoneda et al.}, 1993; {\em Bradley}, 2003). 
Observations show that there is no 70~$\mu$m emission, and this indicates that the dust is warm, 
originating from dust grains with a low long-wavelength emissivity, i.e. with sizes 
$\lesssim$70~$\mu$m/2$\pi$ $\sim$ 10~$\mu$m, located within a few AU of the star, with the strong 
solid state features arising from a component of small, possibly submicron grains ({\em Beichman et al.}, 2005b).
Upper (3-$\sigma$) limits to the 70~$\mu$m emission (L$_{dust}$/L$_{star}$ $<$ 5$\times$10$^{-6}$) 
suggest a potential Kuiper Belt less than 5 times as massive as the Solar System's. 
The emission between the crystalline silicate features at 9--11~$\mu$m, 19~$\mu$m and 
23.8~$\mu$m indicates that there is a source of continuum opacity, possibly a small component 
of larger grains ({\em Beichman et al.}, 2005b). The emitting surface area of the dust is large 
(2.7$\times$10$^{23}$ cm$^2$, $>$1000 times the zodiacal 
emission), and the collisional and P-R drag time for submicron (0.25~$\mu$m) 
grains is $<$1000 yr. This indicates that the dust is either produced by the grinding down of a 
dense asteroid belt (22--64 times more massive than the Solar System's) 
located closer to the star, or originates in a transient event. 
{\em Wyatt et al.} (2006) ruled out the massive asteroid belt scenario and suggested that it is
a transient event, likely the result of recent collisions produced when planetesimals 
located in the outer regions were scattered toward the star in a Late Heavy Bombardment-type event. 

\begin{figure*}
 \epsscale{1.2}
 \plotone{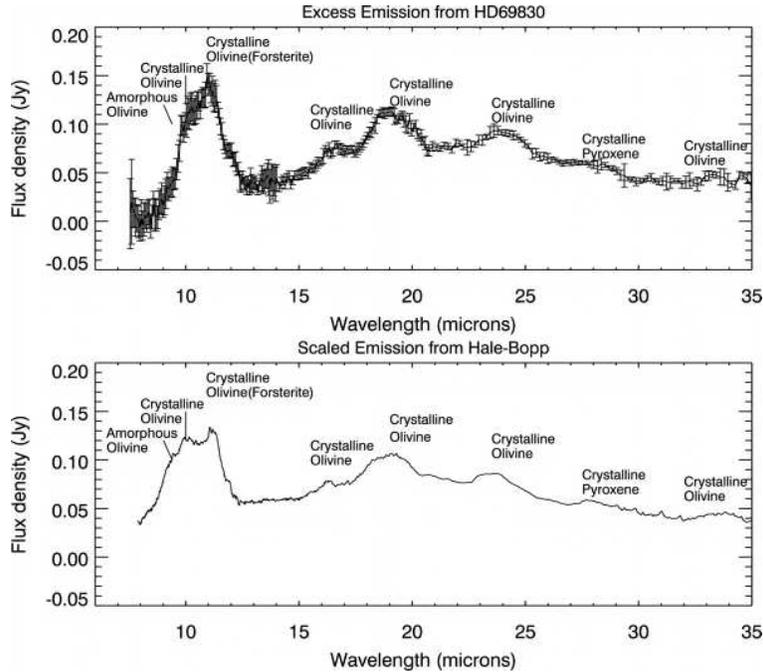}
 \caption{\small Top: Spectrum of the excess of HD 69830. Bottom: spectrum of the comet Hale-Bopp from 
{\em Crovisier et al.} (1996) normalized to a blackbody temperature of 400 K to ease the comparison 
of the two spectra (the observed blackbody temperature is 207 K). Figure from {\em Beichman et al.} (2005b). 
}
\end{figure*}

The disk around $\beta$-Pic seems to be ``normal'' in terms of its mass
content with respect to the stellar age, and does not contain large amounts of 
small silicate grains; on the other hand, the disks around HIP 8920 and HD 69830 are unusually 
dusty and show strong silicate emission features, indicating that silicate features may 
be related to recent collisional events ({\it Weinberger et al.}, priv. comm.). 

The composition of the disk can also be studied from the colors of the scattered 
light images. In general, debris disks are found to be red or neutral. Their redness has
commonly been explained by the presence of 0.4~$\mu$m silicate grains, but except for the two
exception mentioned above (HIP 8920 and HD 69830), spatially resolved spectra have shown 
that debris disks do not generally contain large amounts of small silicate grains; 
a possible explanation for the colors could be that grains are intrinsically red, 
perhaps due to an important contribution from organic materials
({\em Weinberger et al.}, priv. comm; see also {\em Meyer et al.}, 2006). 
For comparison, KBOs present a wide range of surface colors, varying from neutral to very red
(see chapter by {\em Doressoundiram et al.} in this book). 

\bigskip

{\textbf{ 5. DEBRIS DISKS AND CLOSE-IN PLANETS:}} 

{\textbf{ ~~~~RELATED PHENOMENA?}}
\bigskip

The observation of debris disks indicates that planetesimal formation
has taken place around other stars. In these systems, did planetesimal 
formation proceed to the formation of one or move massive planets, as 
was the case of the Sun? In the following cases the answer is yes: 
HD 33636, HD 50554, HD 52265, HD 82943, HD 117176 and HD 128311 are
stars known from radial velocity observations to have at least one planet, 
and they all show 70 $\mu$m~excess (with an excess SNR of 15.4, 14.9, 4.3, 
17.0, 10.2 and 7.1, respectively) arising from cool 
material (T$<$100 K) located mainly beyond 10 AU, implying the presence
of outer belt of dust-producing plantesimals. Their fractional luminosities, 
L$_{dust}$/L$_{star}$, in the range (0.1--1.2)$\times$10$^{-4}$
are $\sim$100 times that inferred for the KB ({\em Beichman et al.}, 2005a). 
Similarly, HD 38529 is a two-planet system that also shows 70 $\mu$m~excess 
emission (with an excess SNR of 4.7; {\em Moro-Mart\'{\i}n et al.}, 2007b). 
HD 69830 is a three-planet system with a strong 24 $\mu$m~excess 
(see $\S$4; {\em Beichman et al.}, 2005b). And finally, $\epsilon$-Eridani has 
at least one close-in planet ({\em Hatzes et al.}, 2000) 
and a spatially resolved debris disk ({\em Greaves et al.}, 2005). 

The nine systems above confirm that debris disks 
and planets co-exist. But are debris disks and the presence of massive planets 
related phenomena? {\em Moro-Mart\'{\i}n et al.} (2007a) found that from the observations of the 
Spitzer Legacy Program FEPS and the GTO results in {\em Bryden et al.} (2006),
there is no sign of correlation between the presence of IR excess 
and the presence of radial velocity planets (see also 
{\em Greaves et al.}, 2004a). This, together with the observation that
high stellar metallicities are correlated with the presence of giants planets
({\em Fischer and Valenti}, 2005) but not correlated with the presence of debris disks 
({\em Greaves, Fischer and Wyatt}, 2006), may indicate that planetary systems with KBOs producing debris 
dust by mutual collisions may be more common than planetary systems harboring gas 
giant planets ({\em Greaves, Fischer and Wyatt}, 2006; {\em Moro-Mart\'{\i}n et al.}, 2007a). 

Most of the debris disks detected with $\it{Spitzer}$ emit only at 70~$\mu$m, i.e. 
the dust is mainly located at distances $>$10 AU, 
while the giant planets detected by radial velocity studies are located within 
a few AU of the star, so the dust and the giant planet(s) could be dynamically
unconnected (but see {\em Moro-Mart\'{\i}n et al.}, 2007b). What about more distant 
giant planets? Do debris disk observations 
contain evidence for long-period planets? We discuss this issue in the next section. 

\bigskip
\noindent
\textbf{ 6. DEBRIS DISK STRUCTURE}
\bigskip

The gravitational perturbations produced by a massive planet on both, the 
dust-producing planetesimals and on the dust particles themselves, 
can create structure in the debris disk giving rise to observable features 
(see e.g. {\em Roques et al.}, 1994; {\em Mouillet et al.}, 1997; {\em Wyatt et al.}, 1999; 
{\em Wyatt}, 2005, 2006;  
{\em Liou and Zook}, 1999; {\em Moro-Mart\'{\i}n and Malhotra}, 2002, 2003, 2005; {\em Moro-Mart\'{\i}n, Wolf and 
Malhotra}, 2005; {\em Kuchner and Holman}, 2003). 

If the disk is radiatively-dominated, $\it{M}$$_{dust}$$\lesssim$10$^{-3}$~\mearth, 
as is the case of the KB dust disk, and if the system contains an outer belt of
planetsimals and one or more inner planets, the disk structure is created because 
the dust grains migrate inward due to the effect of P-R drag, eventually coming in 
resonance with the planet and/or crossing its orbit. This has important consequences 
on their dynamical evolution and therefore on the debris disk structure. 

If the disk is collisionally-dominated, 
$\it{M}$$_{dust}$$\gtrsim$10$^{-3}$~\mearth, before the dust grains migrate far from
their parent bodies, they will suffer frequent collisions that could grind them 
down into smaller grains that are blown away by radiation pressure. 
In this case, the dust grains may not survive
long enough to come into resonance with an inner planet. However, the structure
of the KBOs gives strong evidence that Neptune migrated outward. This process
may have also taken place in other planetary systems, where the outward migration
of a planet could have scattered planetesimals out of the system or trapped 
them into Plutino-like orbits. Because the larger dust particles trace the location
of the parent bodies, this outward migration can strongly affect the debris 
disk structure. 

In this section we summarize the processes by which planets can affect the 
debris disk structure and the observational evidence that indicates that
planets may be responsible for some of the features seen. 

\bigskip
\noindent
\textbf{6.1 Theoretical Predictions}

\bigskip
\noindent
{\em 6.1.1. Gravitational Scattering}\\
\\
Massive planets can eject planetesimals and dust particles out of the planetary system via gravitational
scattering. In the radiatively-dominated disks, if the sources of dust are outside the orbit of the planet, 
this results in an $\it{inner~cavity}$, 
a lower density of dust within the planet's orbit, as the particles drifting inward 
due to P-R drag are likely to be scattered out of the system when crossing the orbit
of the planet ({\em Roques et al.}, 1994). Similarly, planetesimals can get scattered out by a planet
migrating outward, resulting in a depletion of planetesimals and dust inside the orbit of the planet.
Planets with masses of 3 M$_{Jup}$--10 M$_{Jup}$ located 
between 1 AU--30 AU in a circular orbit around a solar type star
eject $>$90\% of the dust grains that go past their orbits by P-R drag; a 
1 M$_{Jup}$ planet at 30 AU ejects $>$80\% of the grains, and about 50\%--90\%
if located at 1 AU, while a 0.3 M$_{Jup}$ planet is not able to open
a gap, ejecting $<$ 10\% of the grains ({\em Moro-Mart\'{\i}n and Malhotra}, 2005). 
These results are valid for dust grains sizes in the range
0.7~$\mu$m--135~$\mu$m, but are probably also applicable to planetesimals (in 
the case of an outward migrating planet), because gravitational scattering
is a process independent of mass as long as the particle under consideration 
can be considered a ``test particle'', i.e. its mass is negligible with respect to that of the planet. 

\bigskip
\noindent
{\em 6.1.2. Resonant Perturbations}\\
\\
Resonant orbits are locations where the orbital period of the planet is $\it{(p+q) / p}$ times 
that of the particle (which can be either a dust grain or a planetesimals), 
where $\it{p}$ and $\it{q}$ are integers, $\it{p}$$>$0 and 
$\it{p+q}$$\geqslant$1. Each resonance has a libration width that depends on the particle 
eccentricity and the planet mass, in which resonant orbits are stable. The 
region close to the planet is chaotic because neighboring resonances overlap
({\em Wisdom}, 1980). Because of the finite width of the resonant region, 
resonant perturbations only affect a small region of the parameter space, but this region 
can be over-populated compared with the size of that parameter space by 
the inward migration of dust particles under the effect of P-R drag
or by the outward migration of the resonance as the planet migrates
({\em Malhotra}, 1993; {\em Malhotra}, 1995; {\em Liou and Zook}, 1995; {\em Wyatt}, 2003). 
When the particle crosses a mean motion resonance ($\it{q}$$>$0), 
it receives energy from the perturbing planet that can balance the energy loss due to 
P-R drag, halting the inward motion of the particle and giving rise to planetary 
resonant rings. Due to the geometry of the resonance, the spatial distribution of material 
in resonance is asymmetric with respect to the planet, being concentrated in clumps. 
There are four basic high-contrast resonant structures that a planet with eccentricity 
$\lesssim 0.6$ can create in a disk of dust released on low eccentricity orbits: a $\it{ring~with~a~gap}$ 
at the location of the planet; a $\it{smooth~ring}$; a $\it{clumpy~eccentric~ring}$; and an $\it{offset~ring}$ 
plus a pair of $\it{clumps}$, with the appearance/dominance of one of these structures depending 
on the mass and eccentricity of the planet ({\em Kuchner and Holman}, 2003).

\bigskip
\noindent
{\em 6.1.3. Secular Perturbations}\\
\\
When a planet is embedded in a debris disk, its gravitational field perturbs the orbits of the particles
(dust grains or planetesimals). 
Secular perturbations are the long-term average of the perturbing forces, and act on timescales $>$0.1 Myr 
(see overview in {\em Wyatt et al.}, 1999). 
As a result of secular perturbations, the planet tries to align the particles with its orbit.
The first particles to be affected are the ones closer to the planet, while the particles
further away are perturbed at a later time, therefore, if the planet's orbital plane is different 
from that of the planetesimal disk, secular perturbations will result in the formation of a $\it{warp}$. 
A warp will also be created if there are two planets on non-coplanar orbits. 

If the planet is in an eccentric orbit, the secular perturbations will force an eccentricity on the 
dust particles, and this will create an $\it{offset}$ in the disk center with respect to the star and 
a $\it{brightness~asymmetry}$ in the re-emitted light, as the dust particles near periastron are 
closer to the star and therefore hotter than the dust particles at the other side of the disk. 

Because secular perturbations act faster on the particles 
closer to the planet, and the forced eccentricities and pericenters are the same for particles located at 
equal distances from the planet, at any one time the secular perturbations of a planet embedded in a 
planetesimal disk can result in the formation of two $\it{spiral~structures}$, one inside and one outside the 
planet's orbit ({\em Wyatt}, 2005b).

\bigskip
\noindent
\textbf{6.2 Observations}
\bigskip

Some of the structural features described above have indeed been observed in the 
spatially resolved images of debris disks (see Figure 5). 

\begin{figure*}
 \epsscale{1.5}
 \plotone{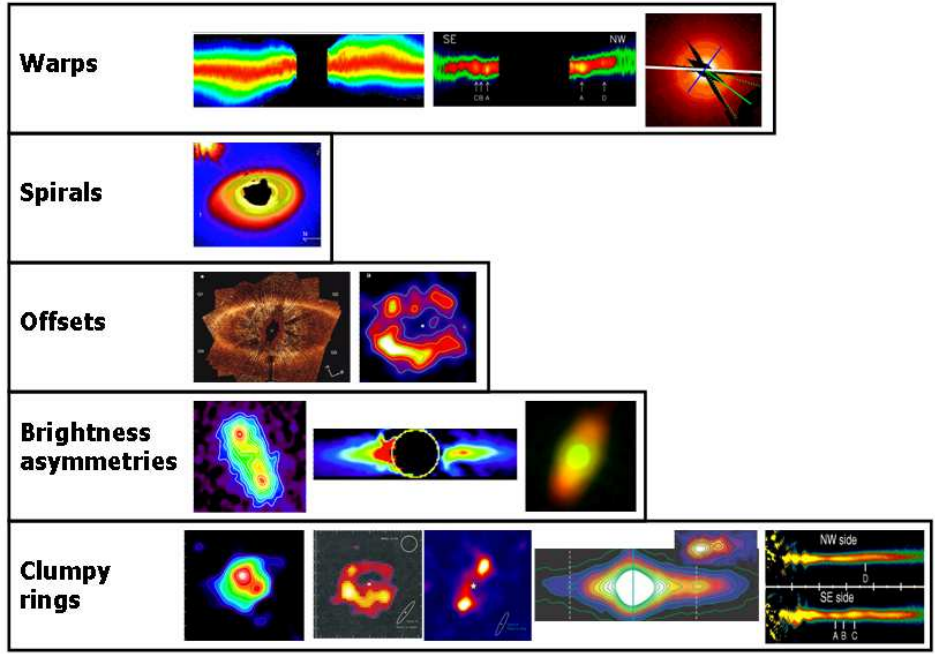}
 \caption{\small Spatially resolved images of debris disks showing a wide diversity of 
debris disk structure. From left to right the images correspond to: 
($\it{1st~row}$) 
$\beta$-Pic (STIS CCD coronography at 0.2--1~$\mu$m; {\em Heap et al.}, 2000), 
AU-Mic (Keck AO at 1.63 ~$\mu$m; {\em Liu}, 2004) and 
TW Hydra (STIS CCD coronography at 0.2--1~$\mu$m; {\em Roberge, Weinberger and Malumuth}, 2005);
($\it{2nd~row}$) HD 141569 (HST/ACS at 0.46--0.72~$\mu$m; {\em Clampin et al.}, 2003); 
($\it{3rd~row}$) Fomalhaut (HST/ACS at 0.69--0.97~$\mu$m; {\em Kalas et al.}, 2005) and 
$\epsilon$-Eri (JCMT/SCUBA at 850~$\mu$m; {\em Greaves et al.}, 2005);
($\it{4th~row}$) HR4796 (Keck/OSCIR at 18.2~$\mu$m; {\em Wyatt et al.}, 1999), 
HD 32297 (HST/NICMOS coronography at 1.1~$\mu$m; {\em Schneider, Silverstone and Hines}, 2005) 
and Fomalhaut (Spitzer/MIPS at 24 and 70~$\mu$m; {\em Stapelfeldt et al.}, 2004); 
($\it{5th~row}$) Vega (JCMT/SCUBA at 850~$\mu$m; {\em Holland et al.}, 1998), 
$\epsilon$-Eri (JCMT/SCUBA at 850~$\mu$m; {\em Greaves et al.}, 1998), 
Fomalhaut (JCMT/SCUBA at 450~$\mu$m; {\em Holland et al.}, 2003),  
$\beta$-Pic (Gemini/T-ReCS at 12.3~$\mu$m; {\em Telesco et al.}, 2005) 
and Au-Mic (HST/ACS at 0.46--0.72~$\mu$m; {\em Krist et al.}, 2005).
All images show emission from 10s to 100s of AU.}
\end{figure*}

\bigskip
\noindent
{\em 6.2.1. Inner Cavities}\\
\\
Inner cavities have long been known to exist. They were first inferred from the $\it{IRAS}$ spectral 
energy distributions (SEDs) of debris disks around A stars, and more recently from the $\it{Spitzer}$ 
SEDs of debris disks around AFGK stars. From the modeling of the disk SED, we can constrain the 
location of the emitting dust by fixing the grain properties. Ideally, the latter can be constrained 
through the modeling of solid state features, however, most debris disk spectroscopy observations 
show little or no features, in which cases it is generally assumed that the grains 
have sizes $\gtrsim$10~$\mu$m and are composed of ``astronomical silicates'' (i.e. silicates with 
optical constants from {\em Weingartner and Draine}, 2001). In most cases, the SEDs show a 
depletion (or complete lack) of mid-infrared thermal emission that is normally associated 
with warm dust located close to the star, and this lack of emission implies the
presence of an inner cavity (or more accurately, a depletion of grains that could be traced 
observationally -- see e.g. {\em Meyer et al.}, 2004; {\em Beichman et al.}, 2005a; {\em Bryden et al.}, 2006; 
{\em Kim et al.}, 2005; {\em Moro-Mart\'{\i}n, Wolf and Malhotra}, 2005; 
{\em Moro-Mart\'{\i}n et al.}, 2007b; {\em Hillenbrand et al.}, in preparation).  

Spatially resolved observations of nearby debris disks have
confirmed the presence of central cavities. From observations in scattered light, 
{\em Kalas et al.} (2006) concluded that debris disks show two basic architectures, 
either narrow belts about 20--30 AU wide and with well-defined outer boundaries (HR 4796A, 
Fomalhaut and HD 139664); or
wide belts with sensitivity-limited edges implying widths $>$50 AU 
(HD 32297, $\beta$-Pic, AU-Mic, HD 107146 and HD 53143). 
Millimeter and sub-millimeter observations show that inner cavities are also present in 
$\epsilon$ Eri (50 AU; {\em Greaves et al.}, 1998), Vega (80 AU; {\em Wilner et al.}, 2002) 
and $\eta$ Corvi (100 AU; {\em Wyatt et al.}, 2005). 

Are all these cavities created by the gravitational ejection of dust 
by massive planets? {\em Wyatt} (2005a) pointed out that because of 
the limited sensitivity of the instruments, most of  the debris disks observed 
so far have large number densities of dust particles and therefore are collisionally-dominated. 
In this regime, mutual collision naturally create inner cavities 
without the need of invoking the presence of a planet 
to scatter out the dust particles.  But this scenario  assumes that 
the parent bodies are depleted from the inner cavity, and the presence of an 
inner edge to the planetesimal distribution may still require the presence 
of a planet. 

Planet formation theories predict the formation of cavities because the planets form 
faster closer to the star, depleting planetesimals from the inner disk regions. 
But planet formation and circumstellar disk evolution are still under debate, so even 
though cavities may be credible evidence for the presence of planets, the connection 
is not well understood. 

\bigskip
\noindent
{\em 6.2.2. Rings and Clumps}\\
\\
Face-on debris disks showing structure that could be associated with 
resonant trapping are Vega ({\em Wilner et al.}, 2002), $\epsilon$-Eridani 
({\em Ozernoy et al.}, 2000; {\em Quillen and Thorndike}, 2002) and Fomalhaut ({\em Wyatt and Dent}, 2002), 
while in edge-on debris disks resonant trapping may lead to the creation 
of brightness asymmetries like those observed in $\beta$-Pic ({\em Thommes and Lissauer}, 2003)
and AU-Mic. 

\bigskip
\noindent
{\em 6.2.3. Warps, Offsets, Spirals and Brightness Asymetries}\\
\\
The debris disk around $\beta$-Pic has two warps, one in the outer disk ({\em Heap et al.}, 2000) 
and another one in the inner disk (with a wavy structure consisting of 4 clumps with counter 
parts at the other side of the disk and none of them aligned with each other; {\em Wahhaj}, 2005). 
New $\it{Hubble}$/ACS observations in scattered light show that the inner ``warp'' in beta-Pic is
really a secondary disk inclined by 5 degrees with respect to the primary disk. This secondary
disk extends to $\sim$80 AU and is probably sustained by a planet that has perturbed planetesimals 
from the outer primary disk into coplanar orbits. Another debris disk showing a warp is AU Mic, 
where the outer part of the disk ($>$80 AU) is tilted by 3 degrees, while the rest of the disk is seen 
mostly edge-on. 

The debris disks around HR 4796 shows a 5\% brightness asymmetry that could be the result 
of a small forced eccentricity imposed by the binary companion HR 4796B, or by an unseen planet located near 
the inner edge of the disk ({\em Wyatt et al.}, 1999). Other debris disks showing brightness asymmetries are
HD 32297 ({\em Schneider, Silverstone and Hines}, 2005) and Fomalhaut ({\em Stapelfeldt et al.}, 2004), and showing offsets are 
Fomalhaut (15 AU; {\em Kalas}, 2005) and $\epsilon$-Eridani (6.6 AU--16.6 AU; {\em Greaves et al.}, 2005). 

A spiral structure has been seen at 325 AU in the debris disk around HD 141569, thought
to be created by a 0.2 M$_{Jup}$--2~M$_{Jup}$ planet located at 235 AU--250 AU with an eccentricity of 
0.05--0.2 ({\em Wyatt}, 2005b).

In summary, dynamical simulations show that gravitational perturbations by a massive planet
can result in the formation of the inner cavities, warps, offsets, brightness asymmetries, spirals, 
rings and and clumps, and these features have indeed been observed in several debris disks. 

\bigskip
\noindent
\textbf{6.3 Other Possible Causes of Debris Disk Structure}
\bigskip

Clumps could trace the location of a planetesimal suffering a recent massive collision, instead
of the location of dust-producing planetesimals or dust particles trapped in mean motion 
resonances with a planet. 
This alternative interpretation has been proposed to explain the brightness asymmetries seen in the mid-IR 
observations of the inner $\beta$-Pic disk ({\em Telesco et al.}, 2005). The brightness asymmetry could arise from 
the presence of a bright clump composed of dust particles with sizes smaller than those in the main disk, 
that could be the result of the collisional grinding of resonantly trapped planetesimals (making the 
clump long-lived, and likely to be observed), or the recent cataclysmic break-up of a planetesimal
with a size $>$100 km (in which case there is no need to have a massive planet in the system, with
the disadvantage that the clump is short-lived and we are observing it at a very particular time, 
maybe within $\sim$50 yr of its break-up; {\em Telesco et al.}, 2005).  However, the clumps seen in the submillimeter 
in systems like Fomalhaut are not easily explained by catastrophic planetesimal collisions because the 
dust masses involved are too large, implying the unlikely collision of two $\sim$ 1400 km sized planetesimals
({\em Wyatt and Dent}, 2002). Brightness asymmetries could also be due to ``sandblasting'' of a debris disk by 
interstellar dust particles, as the star moves with respect to the ISM, but this effect would only affect 
(if anything) the outskirts of the disk, $\gtrsim$400 AU from the central star ({\em Artymowicz and Clampin}, 1997). 
Asymmetries and spiral structure can also be produced by binary companions, but e.g. 
cannot explain all structure seen in the HD141569 disk.  And spiral structure and subsequent 
collapse into nested eccentric rings can also be produced by a close stellar flyby ({\em Kalas, Deltorn and Larwood}, 2001). 
This could in principle explain the clumps seen in the NE of the $\beta$-Pic disk, however, it would require
a flyby on the scale of $<$1000 AU and these encounters are expected to be very rare. In addition, now 
the same type of structure is seen in AU-Mic, another star of the same stellar group, making it 
unlikely that both stars suffered such a fine tuned close encounter. 
Other effects that could be responsible for some of the disk features include instrumental artifacts, 
background/foreground objects, dust migration in a gas disk, photoevaporation, interaction with the 
stellar wind and magnetic field, and dust avalanches ({\em Grigorieva, Artymowicz and Thebault}, 2006). 


\bigskip

\bigskip
\noindent
\textbf{ 6.4 Debris Disks as a Planet Detection Technique}

\bigskip

The two well established planet detection techniques are the radial velocity and the transit studies, 
and both are sensitive to close-in planets only. Direct detection of massive
planets has proven to be very difficult even in their younger (i.e. brighter) stages. This means that old long-period
planets are likely to remain elusive in the foreseeable future. However, we have seen that debris disk structure 
is sensitive to the presence of massive planets with a wide range of semimajor axis (out to 100s of AU), 
complementing the parameter space covered by the other techniques. In this regard, the study of debris disk 
structure has the potential to characterize the diversity of planetary systems and to set
constrains on the outward migration of extra-solar "Neptunes". 

However, before claiming that a planet is present in a debris disk system, the models should be 
able to explain observations at different wavelengths and to account for dust particles of different 
sizes. Different wavelengths trace different particles sizes, and different particle sizes 
have different dynamical evolutions that result in different features.
Large particles dominate the emission at longer 
wavelengths, and their location might resemble that of the dust-producing planetesimals. 
The small grains dominate at short wavelengths; they interact with the stellar radiation 
field more strongly so that their lifetime in the disk is shorter, and therefore their 
presence may signal a recent dust-producing event (like a planetesimal collision). 
And even shorter wavelengths are needed to study the warm dust produced by asteroid-like bodies 
in the terrestrial planet region.
In addition, some of the dynamical models are able to make testable predictions, as for example
the position of resonant structures in multi-epoch imaging, as it is expected that they will orbit the 
planet with periods short enough to result in detectable changes within a decade. This rotation may 
have already been detected in $\epsilon$-Eri to a 2-$\sigma$ level ({\em Greaves et al.}, 2006). Dynamical 
models can also predict the location of the planets, but detecting the planet directly is not feasible 
with current technology. 

\bigskip

\centerline{\textbf{7. THE SOLAR SYSTEM DEBRIS DISK}}
\bigskip

Our Solar System harbors a debris disk, and the inner region is known as the zodiacal cloud.
The sources of dust are very heterogeneous: asteroids and comets 
in the inner region, and KBOs and interstellar dust
in the outer region. The relative contributions of each of these 
sources to the dust cloud is likely to have changed with time, and even the present
relative contributions are controversial: from the He content of the interplanetary dust
particles collected at Earth, it is possible to distinguish between low and high velocity 
grains, associated with an asteroidal and a cometary origin, respectively. The ratio between 
the two populations is not well known, but is thought to differ by less than 
a factor of 10. The contribution of the asteroids to the zodiacal cloud is confirmed
by the observation of dust bands (associated with the formation of individual asteroidal 
families), and must amount to at least a few 10\%. The contribution from the comets is also 
confirmed by the presence of dust trails and tails.
In the outer Solar System, on spatial scales that are more relevant for comparison with other 
debris disks, significant dust production is expected from the 
mutual collisions of KBOs and collisions with interstellar grains ({\em Backman and Paresce}, 1993; 
{\em Stern}, 1996; {\em Yamamoto and Mukai}, 1998). There is evidence for the presence of KB dust from the Pioneer 10 
and 11 dust collision events that took place beyond the orbit of Saturn ({\em Landgraf et al.}, 2002), 
but the dust production rates are still uncertain. 

In parallel to the debris disks properties described in the previous sections, we will now
review some of the properties of the Solar System debris disk. Comparison of these with
the extra-solar systems, can shed some light into the question of whether or not 
our Solar System is unique. 

\bigskip
\noindent
\textbf{ 7.1 Evolution}
\bigskip

Debris disks evolve with time. Therefore, 
the imaging of debris disks at different evolutionary stages could be equivalent to a Solar 
System ``time machine''. However, one needs to be cautious when comparing different systems
because: 
(1) the initial conditions and forming environment of the disks may be significantly different (see $\S$2.1); 
(2) the Solar System debris disk is radiatively-dominated, while the extra-solar debris disks observed
so far, being significantly more massive, are collisionally-dominated, so they are in different physical regimes; and 
(3) the physical processes affecting the later evolution of the disks depend strongly on the planetary configuration, 
e.g. by exciting and/or ejecting planetesimals, and radial velocity observations indicate that planetary configurations 
are very diverse. 
With those caveats in mind, we can 
draw some broad similarities between the time evolution of debris disks and the dust 
in our Solar System. 

As we saw in $\S$3, debris disk evolution consists of a slow decay of dust mass, punctuated by spikes of
high activity, possibly associated with stochastic collisional events. 
Similarly, numerical simulations by {\em Grogan et al.} (2001) indicated that over the lifetime
of the Solar System, the asteroidal dust surface area slowly declined by a factor of 10, and that 
superimposed on this slow decay, asteroidal collisions produced sudden increases of up to an 
order of magnitude, with a decay time of several Myr. Overall, for the 4 Gyr old Sun, the dust 
surface area of the zodiacal cloud is about twice its quiescent level for 10\% of the time. 
Examples of stochastic events in the recent Solar System history are the fragmentation of the asteroid giving
rise to the Hirayama asteroid families, the creation 8.3 Myr ago of the Veritas 
asteroid families, that gave rise to a collisional cascade still accounting for $\sim$25\% of the
zodiacal thermal emission ({\em Dermott et al.}, 2002), as well as collisional 
events resulting in the formation of the dust bands observed by $\it{IRAS}$ ({\em Sykes and Greenberg}, 1986). 
In addition to these small ``spikes'' in the dust production rate at late times, 
there has been one major event in the early Solar System evolution that produced much larger quantities of dust.
Between 4.5 Gyr to 3.85 Gyr there was a heavy cratering phase 
that resurfaced the Moon and the terrestrial planets, creating the lunar basins and 
leaving numerous impact craters in the Moon, Mercury and Mars (all with 
little surface erosion). This ``Heavy Bombardment'' ended $\sim$3.85 Gyr ago, 
600 Myr after the formation of the Sun. Thereafter, the impact rate decreased exponentially with
a time constant ranging from 10 Myr--100 Myr ({\em Chyba}, 1990). 
{\em Strom et al.} (2005) argue that the impact crater record of the terrestrial planets
show that the Late Heavy Bombardment was an event lasting 20--200 Myr, that the source of the impactors 
was the main asteroid belt, and that the mechanism for this event was 
the orbital migration of the giant planets which caused a resonance 
sweeping of the asteroid belt and a large scale ejection of asteroids into planet-crossing
orbits. This event would have been accompanied by a high rate of asteroid 
collisions; the corresponding high rate of dust production would have caused
a large spike in the warm dust luminosity of the Solar System. Although this phenomenon
has not been modeled in any detail, it is likely to be similar to the spikes 
inferred for extra-solar debris disks. 

A massive clearing of planetesimals is thought to have occurred also in the Kuiper Belt. 
This is inferred from the observation that the total mass in the KB region (30 AU--55 AU) is 
$\sim$0.1 \mearth, insufficient to have been able to form the KBOs within the age 
of the Solar System ({\em Stern}, 1996). It is estimated that the primordial KB had a mass
of 30 \mearth--50 \mearth~between 30 AU-55 AU, and was heavily depleted after Neptune formed and
started to migrate outward ({\em Malhotra, Duncan and Levison}, 2000; {\em Levison et al.}, 2006). 
This resulted in the clearing of KBOs with perihelion distances
near or inside the orbit of Neptune, and in the excitation of the KBOs' orbits, which increased 
their relative velocities from 10s m/s to $>$1~km/s,  making their collisions violent enough 
to result in a significant mass of the KBOs ground down to dust and blown away by radiation 
pressure. 

As we have seen in $\S$3.2.2 and $\S$4, detailed studies of nearby debris disks show that 
unusually high dust production rates are needed to explain the properties of several stars, 
including Vega, $\zeta$-Lep, HIP 8920, HD 69830 and $\eta$ Corvi.   
Even though one needs to be cautious about claiming that we are observing all 
these stars at a very special time during their evolution (possibly equivalent to the Late 
Heavy Bombardment), this remains to date the most straightforward explanation of their 
``unusual'' properties. 

Observations therefore indicate that the solar and extra-solar debris disks may have evolved in 
broadly similar ways, in the sense that their dust production decays with time but is 
punctuated by short periods of increased dust production. However, the details of this 
evolution and the comparison of the absolute quantities of dust produced are difficult to assess. 
Preliminary results from the $\it{Spitzer}$ FGK survey ({\em Bryden et al.}, 2006) indicated that even
though the disks observed have a luminosity of $\sim$100 times that of the KB dust disk, 
using the observed cumulative distribution and assuming the distribution of disk luminosities
follows a Gaussian distribution, the observations are consistent with the Solar System having
an order of magnitude greater or less dust than the typical level of dust found around similar 
nearby stars, with the results being inconsistent with most stars having disks much brighter 
than the Solar System's. However, from the $\it{Spitzer}$ $\it{FEPS}$ Legacy, {\em Meyer et al.} 
(2006) arrives to a different preliminary conclusion, suggesting that at times before the 
Late Heavy Bombardment (10 Myr--300 Myr), the dust production rate in the Solar System was much 
higher than that found around stars of similar ages, while at times after the Late Heavy Bombardment 
(1 Gyr--3 Gyr), the dust production rate was much lower than average. 
For example, $\tau$-Ceti is a G8V (solar-type) star with an estimated age of 10 Gyr, surrounded by a 
debris disk that is 20 times dustier than the Solar System's Kuiper Belt ({\em Greaves et al.}, 2004b). 
Which star is ``normal'', $\tau$-Ceti or the Sun? If the present dust production rate in 
$\tau$-Ceti has been going on for the last 10 Gyr, shouldn't all these dust-producing 
planetesimals have been ground down to dust? Have potential planets around $\tau$-Ceti 
undergone a heavy bombardment for the last 10 Gyr, or is the dust the result of a recent 
massive collision? 

\bigskip
\noindent
\textbf{ 7.2 Grain Size and Composition}
\bigskip

As discussed in $\S$4, most debris disk spectra show little or no solid state features, 
indicating that dust particles have grown to sizes $\gtrsim$10~$\mu$m. 
The lack of silicate features, resulting from a lack of small dust grains, 
is also confirmed by the spatially resolved spectroscopy observations of a few nearby debris 
disks. In this regard, our zodiacal cloud is similar to most debris
disks, presenting a predominantly featureless spectrum, thought to arise from dust grains 
10~$\mu$m--100~$\mu$m in size, with a small component of small silicate grains yielding a weak 
(10\% over the continuum) 10~$\mu$m emission feature ({\em Reach et al.}, 2003). 
The analysis of the impact craters on the Long Duration Exposure Facility 
indicated that the mass distribution of the zodiacal dust peaks at 
$\sim$200~$\mu$m ({\em Love and Brownlee}, 1993). The reason why large dust grains are dominant
is a direct result from P-R drag because smaller grains evolve more quickly and therefore
are removed on shorter timescales than larger grains. 
However, for the Solar System, we only have information from the zodiacal cloud, i.e. the warmer 
component of the Solar System's debris disks, because the emission from the colder KB dust 
component is hidden by the inner cloud foreground.

In $\S$4, we also mentioned that there seems to be a correlation between the presence of 
silicate features and large quantities of dust (due possibly to a recent dust-producing event). 
The Solar System, in its quiescent state, seems to be similar (in their lack of small 
silicate grains) to other debris disks that contain ``normal'' amounts of dust for their ages. 
But the Solar System went through periods of high activity, like the Late Heavy Bombardment,
where dust production was orders of magnitude higher. Even though we do not know how the 
Solar System looked like during those spikes in dust production, the remarkable 
similarity between the spectra of the dusty disk around HD~69830 (a 2 Gyr solar type star) and 
comet C/1995 O1 (Hale-Bopp) ({\em Beichman et al.}, 2005b), may indicate that during those stages,
the Solar System's dust disk could have also been similar to other debris disks experiencing 
similar spikes in their dust production. 

\bigskip
\noindent
\textbf{ 7.3 Structure}

\bigskip

The Solar System, being filled with interplanetary dust and harboring planets, is an ideal 
case to investigate the effect of the planets on the dynamics of the 
dust particles, and consequently on the structure of the debris disks. 
Dynamical models predict that the KB dust disk has a density enhancement in a ring-like structure 
between 35--50 AU, with some azimuthal variation due to the trapping into mean motion resonances 
with Neptune and the tendency
of the trapped particles to avoid the resonance planet, creating a minimun density at Neptune's position
({\em  Liou and Zook}, 1999; {\em Moro-Mart\'{\i}n and Malhotra}, 2002; 
{\em Holmes et al.}, 2003; see chapter by {\em Liou and Kaufmann} in this book). 
The models also predict a depletion 
of dust inside 10 AU, due to  gravitational scattering of dust particles by Jupiter and Saturn. 
However, the presence of this structure has not yet been observed (but there is clear evidence 
of the trapping of KBOs in resonance with Neptune, {\em Malhotra}, 1995; {\em Jewitt}, 1999; 
{\em Elliot et al.}, 2005). 

As we mentioned above, the thermal emission from the colder KB dust is hidden by the much 
brighter inner zodiacal cloud foreground, which has been studied in detail by the $\it{IRAS}$, 
$\it{COBE}$ and $\it{ISO}$ space telescopes (that could also map the spatial 
structure of the cloud, as their observing geometry changed throughout the year). These observations, 
together with numerical simulations, revealed that the Earth is embedded in an resonant circumsolar ring 
of asteroidal dust, with a 10\% number density enhancement located in the Earth's wake, giving rise
to the asymetry observed in the zodiacal emission ({\em Jackson and Zook}, 1989; 
{\em Dermott et al.}, 1994; {\em Reach et al.}, 1995). 
In addition, it was found that zodiacal cloud has a warp, as the plane of symmetry of the cloud 
depends on heliocentric distance ({\em Wyatt et al.}, 1999). This ring, the brightness asymmetry and 
the warp, indicate that even though the Solar System debris disk is radiatively-dominated, while the 
extra-solar debris disks observed so far are collisionally-dominated, 
there are some structural features that are common to both.

In terms of disk size, the comparison of the Solar System's dust disk with the handful of nearby spatially 
resolved debris disks observed to date indicates that the Solar System is small. 
This would be consistent with the Sun being born in an OB association, while  kinematic 
studies show that most of the nearby spatially resolved debris disks formed in loosely populated Taurus-like 
associations (see discussion in $\S$2.1). However, it may also be the result of an observational 
bias because so far we have been able to study large disks only. We have to wait until the next generation 
of interferometers come on line to be able to tell whether or not our Solar System debris disk is normal 
in its size. 

\bigskip

\centerline{\textbf{ 8. FUTURE PROSPECTS}}

\bigskip

Debris disks are evidence that many stars are surrounded by dust-producing 
planetesimals, like the asteroids and KBOs in our Solar System. 
In some cases, they also provide evidence of the presence of larger bodies:
first, because the production of dust requires the stirring of planetesimals, 
and the minimum mass for an object needed to start a collisional cascade is 
the mass of Pluto (see chapter by {\em Kenyon et al.} in this book); 
and second, because some debris disk 
show structural features that may be the result of gravitational perturbations
by a Neptune to Jupiter-mass planet. 

Due to limits in sensitivity, we are not yet able to detect debris disks with masses 
similar to that of our Solar System, but only those that are $>$100 times more
massive. Observations are beginning to indicate that the solar and extra-solar 
debris disks may have evolved in broadly similar ways, in the sense that 
their dust production decays with time but is punctuated by short periods of 
increased dust production, possibly equivalent to the Late Heavy Bombardment. 
This offers a unique opportunity to use 
extra-solar debris disks to shed some light in how the Solar System might have 
looked in the past. Similarly, our knowledge of the Solar System is influencing our
understanding of the types of processes which might be at play in the extra-solar 
debris disks. In the future, telescopes like $\it{ALMA}$, $\it{LBT}$, $\it{JWST}$, $\it{TPF}$ and 
$\it{SAFIR}$ will be able to image the dust in planetary systems analogous to our own. 
This will allow to carry out large unbiased surveys sensitive down to the level of dust 
found in our own Solar System that will answer the question of whether or not 
our Solar System debris disk is common or rare.
But very little information is known directly about the KB dust disk, in terms of its mass, 
its spatial structure and its composition, mainly because its thermal emission is overwhelmed 
by the much stronger signal from the inner zodiacal cloud.  
Any advance in understanding of the structure and evolution of the KB is
directly relevant to our understanding of extra-solar
planetary systems. And to that end, there is the need to carry out 
dust experiments on spacecraft traveling to the outer Solar System, like the one onboard $\it{New~Horizons}$, 
and to perform careful modeling of the dynamical evolution of KB dust particles 
and their contribution to the Solar System debris disk that takes into account our increased 
knowledge of the KBOs.
\\
\\
\textbf{ Acknowledgments.} 
A.M.M. is under contract with the Jet Propulsion Laboratory (JPL) funded by NASA through
the Michelson Fellowship Program. JPL is managed for NASA by the California Institute of
Technology. A.M.M. is also supported by the Lyman Spitzer Fellowship at Princeton University. R.M. acknowledges support from NASA-Origins of Solar Systems and NASA-Outer Planets
Research Programs. 

\bigskip

\centerline\textbf{ REFERENCES}
\bigskip
\parskip=0pt
{\small
\baselineskip=11pt

\refs Andrews S. M. and Williams J. P. (2005)
Circumstellar Dust Disks in Taurus-Auriga: The Submillimeter Perspective, 
{\it Astrophys. J., 631}, 1134-1160.

\refs Artymowicz P. and Clampin M. (1997)	
Dust around Main-Sequence Stars: Nature or Nurture by the Interstellar Medium?, 
{\it Astrophys. J., 490}, 863-878.

\refs Aumann H. H., Beichman C. A., Gillett F. C., de Jong T., Houck J. R. et al. (1984)
Discovery of a shell around Alpha Lyrae
{\it Astrophys. J., 278}, L23-L27.

\refs Backman D. E. and Paresce F. (1993) 
Main-sequence stars with circumstellar solid material - The VEGA phenomenon, 
In {\it Protostars and Planets III}, (E.H. Levy and J.I Lunine, eds.), pp. 1253-1304, Univ. of Arizona, Tucson. 

\refs Beichman C. A., Bryden G., Rieke G. H., Stansberry J. A., Trilling D. E. et al. (2005a)
Planets and Infrared Excesses: Preliminary Results from a Spitzer MIPS Survey of Solar-Type Stars, 
{\it Astrophys. J., 622}, 1160-1170. 

\refs Beichman C. A., Bryden G., Gautier T. N., Stapelfeldt K. R., Werner M. W. et al. (2005b)
An Excess Due to Small Grains around the Nearby K0 V Star HD 69830: Asteroid or Cometary Debris?
{\it Astrophys. J., 626}, 1061-1069.

\refs Beichman C. A., Tanner, A., Bryden G., Stapelfeldt K. R., Gautier T. N. (2006a)
IRS Spectra of Solar-Type Stars: A Search for Asteroid Belt Analogs, 
{\it Astrophys. J., 639}, 1166-1176.

\refs Beichman C. A., Bryden G., Stapelfeldt K. R., Gautier T. N., 
Grogan K. et al. (2006b)
New Debris Disks around Nearby Main-Sequence Stars: Impact on the Direct Detection of Planets,
{\it Astrophys. J., 652}, 1674-1693.

\refs Bradley J. (2003)
The Astromineralogy of Interplanetary Dust Particles, 
In {\it Astromineralogy}, (T.K. Henning, ed.), {\it Lecture Notes in Physics, 609}, 217-235.

\refs Bryden G., Beichman C. A., Trilling D. E., Rieke G. H., Holmes E. K. et al. (2006)
Frequency of Debris Disks around Solar-Type Stars: First Results from a Spitzer MIPS Survey, 
{\it Astrophys. J., 636}, 1098-1113. 

\refs Burns J. A., Lamy P. L. and Soter, S. (1979)	
Radiation forces on small particles in the solar system, 
{\it Icarus, 40}, 1-48.

\refs Cameron A. G. W. and Truran J. W (1977)
The supernova trigger for formation of the solar system, 
{\it Icarus, 30}, 447-461.

\refs Chen C. H., Jura M., Gordon K. D. and Blaylock, M. (2005)
A Spitzer Study of Dusty Disks in the Scorpius-Centaurus OB Association, 
{\it Astrophys. J., 623}, 493-501.

\refs Chyba C. F. (1990)
Impact delivery and erosion of planetary oceans in the early inner solar system, 	
{\it Nature, 343}, 129-133. 

\refs Cieza L. A., Cochran W. D. and Paulson, D. B. (2005)
Spitzer Observations of the Hyades: Circumstellar Debris Disks at 625 Myrs Age, 
{\it LPI Contribution No. 1286}, 8421. 

\refs Clampin M., Krist J. E., Ardila D. R., Golimowski D. A., Hartig G. F. et al. (2003)
Hubble Space Telescope ACS Coronagraphic Imaging of the Circumstellar Disk around HD 141569A, 
{\it Astron. J., 126}, 385-392. 

\refs Crovisier J., Brooke T. Y., Hanner M. S., Keller H. U., Lamy P. L. et al. (1996)
Spitzer Observations of the Hyades: Circumstellar Debris Disks at 625 Myrs Age, 
{\it Astron. Astrophys., 315}, L385-L388.

\refs Dermott S. F., Jayaraman S., Xu Y. L., Gustafson B. A. S. and  Liou, J. C. (1994)
A circumsolar ring of asteroidal dust in resonant lock with the Earth, 
{\it Nature, 369}, 719-723. 

\refs Dermott S. F., Kehoe T. J. J., Durda D. D., Grogan K. and Nesvorny, D. (2002)
Recent rubble-pile origin of asteroidal solar system dust bands and asteroidal interplanetary dust particles,
In {\it Proceedings of Asteroids, Comets, Meteors - ACM 2002}, (B. Warmbein, ed.) ESA Publications Division, Noordwijk, 
Netherlands, pp. 319 - 322.

\refs Dominik C. and Decin G. (2003)
Age Dependence of the Vega Phenomenon: Theory, 
{\it Astrophys. J., 598}, 626-635. 

\refs Doressoundiram A., Boehnhardt H., Tegler S. and Trujillo, C. (2007)
Colors, properties and trends of the trans-Neptunian objects,
In {\it Kuiper Belt}, (A. Barucci, H. Boehnhardt, D. Cruikshank, A. Morbidelli, eds.), 
Univ. of Arizona, Tucson. 

\refs Elliot J. L., Kern S. D., Clancy K. B., Gulbis A. A. S., Millis R. L. et al. (2005)
The Deep Ecliptic Survey: A Search for Kuiper Belt Objects and Centaurs. II. Dynamical Classification, the Kuiper Belt Plane, and the Core Population, 
{\it Astron. J., 129}, 1117-1162.

\refs Fischer D. A. and Valenti J. (2005)
The Planet-Metallicity Correlation, 
{\it Astrophys. J., 622}, 1102-1117. 

\refs Gorlova N., Padgett D. L., Rieke G. H., Muzerolle J., Stauffer J. R. et al. (2004)
New Debris-Disk Candidates: 24 Micron Stellar Excesses at 100 Million years, 
{\it Astrophys. J. Supp., 154}, 448-452. 

\refs Gorlova N., Rieke G. H., Muzerolle J., Stauffer J. R., Siegler N. et al. (2006)
Spitzer 24 micron Survey of Debris Disks in the Pleiades
{\it Astrophys. J., 649}, 1028-1042.

\refs Greaves J. S., Holland, W. S., Moriarty-Schieven G., Jenness, T., Dent, W. R. F. et al. (1998)
A Dust Ring around epsilon Eridani: Analog to the Young Solar System, 
{\it Astrophys. J., 506}, L133-L137.

\refs Greaves J. S., Holland W. S., Jayawardhana R., Wyatt M. C. and Dent, W. R. F. (2004a)
A search for debris discs around stars with giant planets, 
{\it Mon. Not. R. Astron. Soc., 348}, 1097-1104.

\refs Greaves J. S., Wyatt M. C., Holland W. S. and Dent, W. R. F. (2004b)
The debris disc around tau Ceti: a massive analogue to the Kuiper Belt, 
{\it Mon. Not. R. Astron. Soc., 351}, L54-L58.

\refs Greaves J. S., Holland W. S., Wyatt M. C., Dent W. R. F., Robson E. I. (2005) 
Structure in the $\epsilon$ Eridani Debris Disk, 
{\it Astrophys. J., 619}, L187-L190. 

\refs Greaves J. S., Fischer D. A. and Wyatt M. C. (2006)
Metallicity, debris discs and planets,
{\it Mon. Not. R. Astron. Soc., 366}, 283-286. 

\refs Grigorieva A., Artymowicz P. and Thebault P. (2006)
Collisional dust avalanches in debris discs, 
{\it Astron. Astrophys., 461}, 537-549. 

\refs Grogan K., Dermott S. F. and Durda, D. D. (2001)
The Size-Frequency Distribution of the Zodiacal Cloud: Evidence from the Solar System Dust Bands, 
{\it Icarus, 152}, 251-267. 

\refs Hartmann L. (2000)
{\it Accretion Processes in Star Formation}, Cambridge University Press.

\refs Hatzes A. P., Cochran W. D., McArthur, B., Baliunas S. L., Walker G. A. H. et al. (2000)
Evidence for a Long-Period Planet Orbiting $\epsilon$-Eridani, 
{\it Astrophys. J., 544}, L145-L148.

\refs Hayashi C. (1981)
Structure of the solar nebula, growth and decay of magnetic fields and effects of magnetic and turbulent viscosities on the nebula, 
{\it Prog. Theor. Phys. Suppl., 70}, 35-53.

\refs Heap S. R., Lindler D. J., Lanz T. M., Cornett R. H., Hubeny I. Maran et al. (2000)
Space Telescope Imaging Spectrograph Coronagraphic Observations of beta Pictoris, 
{\it Astrophys. J., 539}, 435-444.

\refs Hillenbrand L. A. and Hartmann L. W. (1998)
A Preliminary Study of the Orion Nebula Cluster Structure and Dynamics, 
{\it Astrophys. J., 492}, 540-553. 

\refs Hines D. C., Backman, D. E., Bouwman, J., Hillenbrand L. A., Carpenter J. M. et al. (2006)
The Formation and Evolution of Planetary Systems (FEPS): Discovery of an Unusual Debris System Associated with HD 12039, 
{\it Astrophys. J., 638}, 1070-1079.

\refs Holland W. S., Greaves J. S., Zuckerman, B., Webb, R. A., McCarthy C. et al. (1998)
Submillimetre images of dusty debris around nearby stars, 
{\it Nature, 392}, 788-790. 

\refs Holland W. S., Greaves J. S., Dent W. R. F., Wyatt M. C., Zuckerman B. et al. (2003)
Submillimeter Observations of an Asymmetric Dust Disk around Fomalhaut, 
{\it Astrophys. J., 582}, 1141-1146. 

\refs Hollenbach D. and Adams F. C. (2004)
Dispersal of Disks Around Young Stars: Constraints on Kuiper Belt Formation, 
In {\it Debris Disks and the Formation of Planets: A Symposium in Memory of Fred Gillett}, {\it ASP Conference Series, 324}, 
(L. Caroff, L. J. Moon, D. Backman, and E. Praton., eds.),  pp. 168. 

\refs Hollenbach D., Gorti U., Meyer M., Kim J. S., Morris P. et al. (2005)
Formation and Evolution of Planetary Systems: Upper Limits to the Gas Mass in HD 105, 
{\it Astrophys. J., 631}, 1180-1190.

\refs Holmes, E. K., Dermott, S. F., Gustafson, B. A. S. and Grogan, K. (2003)
Resonant Structure in the Kuiper Disk: An Asymmetric Plutino Disk, 
{\it Astrophys. J., 597}, 1211-1236.

\refs Jackson, A. A.; Zook, H. A. (1989)
A solar system dust ring with earth as its shepherd, 
{\it Nature, 337}, 629-631. 

\refs Jewitt D. (1999)
Kuiper Belt Objects, 
{\it Ann. Rev. Earth Planet. Sci., 27}, 287-312.

\refs Jura M., Chen C. H., Furlan E., Green J., Sargent B. et al. (2004)
Mid-Infrared Spectra of Dust Debris around Main-Sequence Stars, 
{\it Astrophys. J. Supp., 154}, 453-457.

\refs Kalas P., Deltorn J.-M. and Larwood, J. (2001)
Stellar Encounters with the beta Pictoris Planetesimal System, 
{\it Astrophys. J., 553}, 410-420. 

\refs Kalas P., Graham J. R. and Clampin M. (2005)
A planetary system as the origin of structure in Fomalhaut's dust belt, 
{\it Nature, 435}, 1067-1070. 

\refs Kalas P., Graham J. R., Clampin M. C. and Fitzgerald M. P. (2006)
First Scattered Light Images of Debris Disks around HD 53143 and HD 139664, 
{\it Astrophys. J., 637}, L57-L60.

\refs Kenyon S. J. and Bromley B. C. (2004)
Collisional Cascades in Planetesimal Disks. II. Embedded Planets, 
{\it Astron. J., 127}, 513-530. 

\refs Kenyon S. J. and Bromley B. C. (2005)
Prospects for Detection of Catastrophic Collisions in Debris Disks, 
{\it Astron. J., 130}, 269-279. 

\refs Kenyon S. J., Bromley B. C., O'Brien, D. P. and Davis, D. R. (2007)
Formation and Collisional Evolution of Kuiper Belt Objects,
In {\it Kuiper Belt}, (A. Barucci, H. Boehnhardt, D. Cruikshank, A. Morbidelli, eds.), 
Univ. of Arizona, Tucson. 

\refs Kim, J. S., Hines D. C., Backman D. E., Hillenbrand, L. A., Meyer, M. R. et al. (2005)
Formation and Evolution of Planetary Systems: Cold Outer Disks Associated with Sun-like Stars, 
{\it Astrophys. J., 632}, 659-669.

\refs Kobayashi H., Ida S. and Tanaka H. (2005) 
The evidence of an early stellar encounter in Edgeworth Kuiper belt
{\it Icarus, 177}, 246-255.

\refs Krist J. E., Ardila D. R., Golimowski D. A., Clampin M., Ford H. C. (2005)
Hubble Space Telescope Advanced Camera for Surveys Coronagraphic Imaging of the AU Microscopii Debris Disk, 
{\it Astron. J., 129}, 1008-1017. 

\refs Kuchner M. J. and Holman M. J. (2003)
The Geometry of Resonant Signatures in Debris Disks with Planets, 
{\it Astrophys. J., 588}, 1110-1120.

\refs Lagrange A.-M., Backman D. E. and Artymowicz P. (2000)
Main-sequence stars with circumstellar solid material - The VEGA phenomenon, 
In {\it Protostars and Planets IV}, (V. Mannings, A.P. Boss, S.S. Russell, eds.), pp. 639, 
Univ. of Arizona, Tucson. 

\refs Landgraf M., Liou J.-C. Zook H. A. and Gr$\ddot{u}$n E (2002)
Origins of Solar System Dust beyond Jupiter, 
{\it Astron. J., 123}, 2857-2861.

\refs Levison H. F., Morbidelli A., Gomes R. and Backman D. (2006)
Planet migration in planetesimal disks, 
In {\it Protostars and Planets V}, (B. Reipurth, D. Jewitt, K. Keil, eds.),
Univ. of Arizona Tucson.

\refs Liou J.-C. and Kaufmann D. A. (2007)
Structure of the Kuiper Belt Dust Disk,
In {\it Kuiper Belt}, (A. Barucci, H. Boehnhardt, D. Cruikshank, A. Morbidelli, eds.), 
Univ. of Arizona, Tucson. 

\refs Liou J.-C. and Zook H. A. (1995)
An asteroidal dust ring of micron-sized particles trapped in 1 : 1 mean motion with Jupiter,
{\it Icarus, 113}, 403-414. 

\refs Liou J.-C. and Zook H. A. (1999)
Signatures of the Giant Planets Imprinted on the Edgeworth-Kuiper Belt Dust Disk, 
{\it Astron. J., 118}, 580-590. 

\refs Liou J.-C., Dermott S. F. and Xu Y. L. (1995)
The contribution of cometary dust to the zodiacal cloud,
{\it Planet. Space Sci., 43}, 717-722. 

\refs Liu M. C. (2004)
Substructure in the Circumstellar Disk Around the Young Star AU Microscopii,
{\it Science, 305}, 1442-1444.

\refs Love S. G. and Brownlee D. E. (1993)
A Direct Measurement of the Terrestrial Mass Accretion Rate of Cosmic Dust, 
{\it Science, 262}, 550-553. 

\refs Malhotra R. (1993)
The Origin of Pluto's Peculiar Orbit, 
{\it Nature, 365}, 819-821. 

\refs Malhotra R. (1995)
The Origin of Pluto's Orbit: Implications for the Solar System Beyond Neptune,
{\it Astron. J., 110}, 420-430. 

\refs Malhotra R., Duncan M. J. and Levison H. F. (2000)
Dynamics of the Kuiper Belt, 
In {\it Protostars and Planets IV}, (V. Mannings, A.P. Boss, S.S. Russell, eds.), pp. 1231, 
Univ. of Arizona, Tucson. 

\refs Meyer M. R., Hillenbrand L. A., Backman D. E., Beckwith S. V. W., Bouwman, J. et al. (2004)
The Formation and Evolution of Planetary Systems: First Results from a Spitzer Legacy Science Program, 
{\it Astrophys. J. Supp., 154}, 422-427.

\refs Meyer M. R., Backman D. E., Weinberger A. J. and Wyatt M. C. (2006)
Evolution of Circumstellar Disks Around Normal Stars: Placing Our Solar System in Context, 
In {\it Protostars and Planets V}, (B. Reipurth, D. Jewitt, K. Keil, eds.),
Univ. of Arizona Tucson.

\refs Moro-Mart\'{\i}n A. and Malhotra R. (2002)
A Study of the Dynamics of Dust from the Kuiper Belt: Spatial Distribution and Spectral Energy Distribution, 
{\it Astron. J., 124}, 2305-2321.  

\refs Moro-Mart\'{\i}n A. and Malhotra R. (2003)
Dynamical Models of Kuiper Belt Dust in the Inner and Outer Solar System, 
{\it Astron. J., 125}, 2255-2265.  

\refs Moro-Mart\'{\i}n A. and Malhotra R. (2005)
Dust Outflows and Inner Gaps Generated by Massive Planets in Debris Disks, 
{\it Astrophys. J., 633}, 1150-1167. 

\refs Moro-Mart\'{\i}n A., Wolf S. and Malhotra R. (2005)
Signatures of Planets in Spatially Unresolved Debris Disks, 
{\it Astrophys. J., 621}, 1079-1097.

\refs Moro-Mart\'{\i}n A., Carpenter J. M., Meyer M. R., Hillenbrand L. A., Malhotra R. et al. (2007a)
Are Debris Disks and Massive Planets Correlated?, 
{\it Astrophys. J., 658}, in press.

\refs Moro-Mart\'{\i}n A., Malhotra R., Carpenter J. M., Hillenbrand L. A., Wolf S. et al. (2007b)
The dust, planetesimals and planets of HD 38529, 
{\it Astrophys. J.}, submitted.

\refs Mouillet D., Larwood J. D., Papaloizou J. C. B. and Lagrange A. M. (1997)
A planet on an inclined orbit as an explanation of the warp in the Beta Pictoris disc, 
{\it Mon. Not. R. Astron. Soc., 292}, 896-904. 

\refs Natta A. (2004)
Circumstellar Disks in Pre-Main Sequence Stars, 
In {\it Debris Disks and the Formation of Planets: A Symposium in Memory of Fred Gillett}, {\it ASP Conference Series, 324}, 
(L. Caroff, L. J. Moon, D. Backman, and E. Praton., eds.),  pp. 20. 

\refs Okamoto Y. K., Kataza H., Honda M., Yamashita, T. et al. (2004)
An early extrasolar planetary system revealed by planetesimal belts in beta Pictoris, 
{\it Nature, 431}, 660-663. 

\refs Ozernoy L. M., Gorkavyi N. N., Mather J. C. and Taidakova T. A. (2000)
Signatures of Exosolar Planets in Dust Debris Disks, 
{\it Astrophys. J., 537}, L147-L151.

\refs Pascucci I., Gorti U., Hollenbach D., Najita J. and Meyer M.R. (2006)
Formation and evolution of planetary systems: upper limitis to the gas mass in disks around solar-like stars. 
{\it Astrophys. J., 651}, 1177-1193.

\refs Quillen A. C. and Thorndike S. (2002)
Structure in the $\epsilon$ Eridani Dusty Disk Caused by Mean Motion Resonances with a 0.3 Eccentricity Planet at Periastron, 
{\it Astrophys. J., 578}, L149-L152.

\refs Reach W. T., Franz B. A., Weiland J. L., Hauser M. G., Kelsall T. N. et al. (1995)
Observational Confirmation of a Circumsolar Dust Ring by the COBE Satellite, 
{\it Nature, 374}, 521-523. 

\refs Reach W. T., Morris P., Boulanger, F. and Okumura K. (2003)
The mid-infrared spectrum of the zodiacal and exozodiacal light, 
{\it Icarus, 164}, 384-403.

\refs Rieke G. H., Su K. Y. L., Stansberry J. A., Trilling D., Bryden G. et al. (2005)
Decay of Planetary Debris Disks, 
{\it Astrophys. J., 620}, 1010-1026.

\refs Roberge A., Weinberger A. J. and Malumuth E. M. (2005)
Spatially Resolved Spectroscopy and Coronagraphic Imaging of the TW Hydrae Circumstellar Disk, 
{\it Astrophys. J., 622}, 1171-1181.

\refs Roques F., Scholl H., Sicardy B. and Smith B. A. (1994)
Is there a planet around beta Pictoris? Perturbations of a planet on a circumstellar dust disk. 1: The numerical model, 
{\it Icarus, 108}, 37-58

\refs Schneider G., Silverstone M. D. and Hines D. C. (2005)
Discovery of a Nearly Edge-on Disk around HD 32297, 
{\it Astrophys. J., 629}, L117-L120. 

\refs Shu F. H., Adams F. C. and Lizano, S. (1987)
Star formation in molecular clouds - Observation and theory
{\it Ann. Rev. Astron. Astrophys., 25}, 23-81.

\refs Siegler N., Muzerolle J., Young E. T., Rieke G. H., Mamajek E. et al. (2006)
Spitzer 24 micron observations of open cluster IC 2391 and debris disk evolution of FGK stars, 
{\it Astrophys. J., 654}, 580-594.

\refs Simon M., Dutrey A. and Guilloteau S. (2000)
Dynamical Masses of T Tauri Stars and Calibration of Pre-Main-Sequence Evolution
{\it Astrophys. J., 545}, 1034-1043.

\refs Song I., Zuckerman B. and Bessell M. S. (2003)
New Members of the TW Hydrae Association, beta Pictoris Moving Group, and Tucana/Horologium Association, 
{\it Astrophys. J., 599}, 342-350.

\refs Song I., Zuckerman B., Weinberger A. J. and Becklin E. E. (2005)
Extreme collisions between planetesimals as the origin of warm dust around a Sun-like star, 
{\it Nature, 436}, 363-365. 

\refs Stapelfeldt, K. R., Holmes E. K., Chen C., Rieke G. H. and Su K. Y. L. et al. (2004)
First Look at the Fomalhaut Debris Disk with the Spitzer Space Telescope, 
{\it Astrophys. J. Supp., 154}, 458-462. 

\refs Stauffer J. R., Rebull, L. M., Carpenter, J., Hillenbrand, L., Backman, D. et al. (2005)
Spitzer Space Telescope Observations of G Dwarfs in the Pleiades: Circumstellar Debris Disks at 100 Myr Age, 
{\it Astron. J., 130}, 1834-1844. 

\refs Stern S. A. (1996)
On the Collisional Environment, Accretion Time Scales, and Architecture of the Massive, Primordial Kuiper Belt,
{\it Astron. J., 112}, 1203-1214.

\refs Strom R. G., Malhotra R., Ito T., Yoshida F. and Kring D. A. (2005)
The Origin of Planetary Impactors in the Inner Solar System, 
{\it Science, 309}, 1847-1850.

\refs Su K. Y. L., Rieke G. H., Stansberry J. A., Bryden G., Stapelfeldt K. R. et al. (2006)
Debris disk evolution around A stars,  
{\it Astrophys. J.}, 653, 675-689.

\refs Sykes M. V. and Greenberg R. (1986)
The formation and origin of the IRAS zodiacal dust bands as a consequence of single collisions between asteroids, 
{\it Icarus, 65}, 51-69.

\refs Tachibana S., Huss G. R., Kita N. T., Shimoda G. and Morishita Y. (2006)
60Fe in Chondrites: Debris from a Nearby Supernova in the Early Solar System?, 
{\it Astrophys. J., 639}, L87-L90.

\refs Telesco C. M., Fisher R. S., Wyatt, M. C., Dermott, S. F., Kehoe T. J. J. et al. (2005) 
Mid-infrared images of beta Pictoris and the possible role of planetesimal collisions in the central disk, 
{\it Nature, 433}, 133-136.

\refs Thommes E. W. and Lissauer J. J. (2003)
Resonant Inclination Excitation of Migrating Giant Planets, 
{\it Astrophys. J., 597}, 566-580. 

\refs Wahhaj, Z. (2005)
Planetary signatures in circumstellar debris disks, 
{\it Ph.D dissertation}. 

\refs Weidenschilling S. J. (1977)
The distribution of mass in the planetary system and solar nebula
{\it Astrophys. Space Sci., 51}, 153-158.

\refs Weinberger A. J., Becklin E. E. and Zuckerman B. (2003)
The First Spatially Resolved Mid-Infrared Spectroscopy of beta Pictoris, 
{\it Astrophys. J., 584}, L33-L37.

\refs Weingartner J. C. and Draine B. T. (2001) 
Dust Grain-Size Distributions and Extinction in the Milky Way, Large Magellanic Cloud, and Small Magellanic Cloud, 
{\it Astrophys. J., 548}, 296-309. 

\refs Wilner D. J., Holman M. J., Kuchner M. J. and Ho P. T. P (2002)
Structure in the Dusty Debris around Vega, 
{\it Astrophys. J., 569}, L115-L119.

\refs Wisdom J. (1980)
The resonance overlap criterion and the onset of stochastic behavior in the restricted three-body problem, 
{\it Astron. J., 85}, 1122-1133. 

\refs Wyatt M. C. (2003)
Resonant Trapping of Planetesimals by Planet Migration: Debris Disk Clumps and Vega's Similarity to the Solar System, 
{\it Astrophys. J., 598}, 1321-1340.

\refs Wyatt M. C. (2005a)
The insignificance of P-R drag in detectable extrasolar planetesimal belts, 
{\it Astron. Astrophys., 433}, 1007-1012.  

\refs Wyatt M. C. (2005b)
Spiral structure when setting up pericentre glow: possible giant planets at hundreds of AU in the HD 141569 disk
{\it Astron. Astrophys., 440}, 937-948. 

\refs Wyatt M. C. (2006)
Dust in Resonant Extrasolar Kuiper Belts: Grain Size and Wavelength Dependence of Disk Structure, 
{\it Astrophys. J., 639}, 1153-1165.

\refs Wyatt M. C. and Dent (2002)
Collisional processes in extrasolar planetesimal discs - dust clumps in Fomalhaut's debris disc, 
{\it Mon. Not. R. Astron. Soc., 334}, 589-607.

\refs Wyatt M. C., Dermott S. F., Telesco C. M., Fisher R. S., Grogan K. et al. (1999)
How Observations of Circumstellar Disk Asymmetries Can Reveal Hidden Planets: Pericenter Glow and Its Application to the HR 4796 Disk, 
{\it Astrophys. J., 527}, 918-944. 

\refs Wyatt M. C., Greaves J. S., Dent W. R. F. and Coulson I. M. (2005)
Submillimeter Images of a Dusty Kuiper Belt around eta Corvi,
{\it Astrophys. J., 620}, 492-500.

\refs Wyatt M. C., Smith R., Greaves J. S., Beichman C. A., Bryden G. and Lisse C. M. (2006) 
Transience of hot dust around sun-like stars,
{\it Astrophys. J., 657}, in press. 

\refs Yamamoto S. and  Mukai T. (1998)
Dust production by impacts of interstellar dust on Edgeworth-Kuiper Belt objects,
{\it Astron. Astrophys., 329}, 785-791.

\refs Yoneda S., Simon S. B., Sylvester P. J., Hsu A. and Grossman L. (1993)
Large Siderophile-Element Fractionations in Murchison Sulfides, 
{\it Meteoritics, 28}, 465-516. 

\end{document}